\def\beq#1{\begin{equation}\label{#1}}
\def\eeq{\end{equation}}
\def\beqa#1{\begin{eqnarray}\label{#1}}
\def\eeqa{\end{eqnarray}}

\def\fun#1#2{\lower3.6pt\vbox{\baselineskip0pt\lineskip.9pt
        \ialign{$\mathsurround=0pt#1\hfill##\hfil$\crcr#2\crcr\sim\crcr}}}

\def\xi{{{\bf x}^b}}

\documentclass
[twocolumn,aps,showpacs,showkeys,nofootinbib,superscriptaddress]{revtex4}
\usepackage{epsfig}

\newcommand{\be}{\begin{equation}}
\newcommand{\ee}{\end{equation}}
\newcommand{\ba}{\begin{eqnarray}}
\newcommand{\ea}{\end{eqnarray}}

\begin{document}
\input{epsf.sty}

\title{Exploring the evolution of color-luminosity parameter $\beta$ and its effects on parameter estimation}

\author{Shuang Wang}
\email{wjysysgj@163.com}
\affiliation{Department of Physics, College of Sciences, Northeastern University, Shenyang 110004, China}

\author{Yun-He Li}
\email{liyh19881206@126.com}
\affiliation{Department of Physics, College of Sciences, Northeastern University, Shenyang 110004, China}

\author{Xin Zhang\footnote{Corresponding author}}
\email{zhangxin@mail.neu.edu.cn}
\affiliation{Department of Physics, College of Sciences, Northeastern University, Shenyang 110004, China}
\affiliation{Center for High Energy Physics, Peking University, Beijing 100080, China}

\begin{abstract}
It has been found in previous studies that, for the Supernova Legacy Survey three-year (SNLS3) data,
there is strong evidence for the redshift-evolution of color-luminosity parameter $\beta$.
In this paper, using three simplest dark energy models ($\Lambda$CDM, $w$CDM, and CPL),
we further explore the evolution of $\beta$ and its effects on parameter estimation.
In addition to the SNLS3 data,
we also take into account the Planck distance priors data,
as well as the latest galaxy clustering (GC) data extracted from SDSS DR7 and BOSS.
We find that, for all the models,
adding a parameter of $\beta$ can reduce $\chi^2_{min}$ by $\sim$ 36,
indicating that $\beta_1 = 0$ is ruled out at 6$\sigma$ confidence levels.
In other words, $\beta$ deviates from a constant at 6$\sigma$ confidence levels.
This conclusion is insensitive to the dark energy models considered,
showing the importance of considering the evolution of $\beta$ in the cosmology-fits.
Furthermore,
it is found that varying $\beta$ can significantly change the fitting results of various cosmological parameters:
using the SNLS3 data alone,
varying $\beta$ yields a larger $\Omega_m$ for the $\Lambda$CDM model;
using the SNLS3+CMB+GC data,
varying $\beta$ yields a larger $\Omega_m$ and a smaller $h$ for all the models.
Moreover, we find that these results are much closer to those given by the CMB+GC data,
compared to the cases of treating $\beta$ as a constant.
This indicates that considering the evolution of $\beta$ is very helpful
for reducing the tension between supernova and other cosmological observations.

\end{abstract}



\maketitle

\section{Introduction}

Various astronomical observations \cite{Riess98,spergel03,Tegmark04,Komatsu09,Percival10,Drinkwater10,Riess11}
all indicate that the Universe is undergoing an accelerated expansion.
So far, we are still in the dark about the nature of this extremely counterintuitive phenomenon;
it may be due to an unknown energy component, i.e., dark energy~\cite{quint,phantom,k,Chaplygin,tachyonic,HDE,hessence,YMC,others1,others2,WangTegmark05,others3},
or a modification of general relativity~\cite{SH,PR,DGP,GB,Galileon,FR,FT,FRT}.
For recent reviews, see, e.g.,~\cite{CST,FTH,Linder,CK,Uzan,Tsujikawa,NO,LLWW,CFPS,YWBook}.

One of the most powerful probes of dark energy is the use of type Ia supernovae (SNe Ia),
which can be used as cosmological standard candles to measure the expansion history of the Universe.
In recent years, several supernova (SN) datasets with hundreds of SNe Ia were released,
such as ``SNLS''~\cite{Astier06}, ``Union''~\cite{Kowalski08}, ``Constitution''~\cite{Hicken09}, ``SDSS''~\cite{Kessler09},
``Union2''~\cite{Amanullah10} and ``Union2.1''~\cite{Suzuki12}.

In 2010, a high quality SN dataset from the first three years
of the Supernova Legacy Survey (SNLS3) was released \cite{Guy10}.
Soon after, Conley et al. presented SNe-only cosmological results
by combining the SNLS3 SNe with various low- to mid-$z$ samples \cite{Conley11},
and Sullivan et al. presented the joint cosmological constraints
by combining the SNLS3 dataset with other cosmological data sets \cite{Sullivan11}.
In~\cite{Conley11} three SNe data sets are presented, depending on different light-curve fitters:
``SALT2'', which consists of 473 SNe Ia;
``SiFTO'', which consists of 468 SNe Ia;
and ``combined'', which consists of 472 SNe Ia.
It should be stressed that the SNLS team treated two important quantities,
stretch-luminosity parameter $\alpha$ and color-luminosity parameter $\beta$ of SNe Ia,
as free model parameters on the same footing as the cosmological parameters,
all to be estimated during the Hubble diagram fitting process
using the covariance matrix that includes {\it both} statistical and systematic errors.

A critical challenge is the control of the systematic uncertainties of SNe Ia.
One of the most important factors is the effect of potential SN evolution,
i.e., the possibility of the evolution of $\alpha$ and $\beta$ with redshift $z$.
The current studies show that $\alpha$ is still consistent with a constant,
but the hints for the evolution of $\beta$ have already been found in \cite{Astier06,Kowalski08,Kessler09,Guy10,Marriner11,Conley11,Sullivan11}.
It should be pointed out that
these papers studied $\beta$'s evolution using bin-by-bin fits,
which were very difficult to make definitive statements because of the correlations between different bins.
In \cite{Mohlabeng}, Mohlabeng and Ralston firstly used a linear parametrization $\beta(z) = \beta_0 + \beta_1 z$ to study the Union2.1 sample,
and found that $\beta$ deviates from a constant at 7$\sigma$ confidence levels.
Moreover, they proved that using the linear parametrization can obtain better results than using bin-by-bin methods.
In \cite{WangWang}, Wang and Wang studied the case of SNLS3 data
using three functional forms: the linear, the quadratic, and the step function fits.
They found that $\beta$ increases significantly with $z$ when the systematic uncertainties of SNLS3 sample are taken into account,
showing that the evolution of $\beta$ is insensitive to the functional form of $\beta(z)$ assumed.

It is clear that a time-varying $\beta$ has significant impact on parameter estimation.
In \cite{WangWang}, using the cubic spline interpolation for a scaled comoving distance $r_p(z)$,
the effects of varying $\beta$ on distance-redshift relation is briefly discussed.
It is also very interesting to study the impact of varying $\beta$ on various cosmological models.
So in this paper,
we study this issue
by considering three simplest dark energy models: $\Lambda$CDM, $w$CDM, and CPL \cite{CPL}.
For comparison, we also take into account the Planck distance priors data \cite{WangWangCMB},
as well as the latest galaxy clustering (GC) data
extracted from SDSS DR7~\cite{ChuangWang12} and BOSS~\cite{Chuang13}.

We describe our method in Sec. II, present our results in Sec. III, and conclude in Sec. IV.

\section{Method}
\label{sec:method}

The comoving distance to an object at redshift $z$ is given by:
\ba
\label{eq:r(z)}
 & &r(z)=cH_0^{-1}\, |\Omega_k|^{-1/2} {\rm sinn}[|\Omega_k|^{1/2}\, \Gamma(z)],\\
 & &\Gamma(z)=\int_0^z\frac{dz'}{E(z')}, \hskip 1cm E(z)=H(z)/H_0 \nonumber
\ea
where ${\rm sinn}(x)=\sin(x)$, $x$, $\sinh(x)$ for $\Omega_k<0$, $\Omega_k=0$, and $\Omega_k>0$ respectively.
The expansion rate of the universe $H(z)$ (i.e., the Hubble parameter) is given by
\be
 H^2(z)= H_0^2 \left[ \Omega_m (1+z)^3 + \Omega_k (1+z)^2 + \Omega_X X(z) \right],
\ee
where $\Omega_m+\Omega_k+\Omega_X=1$.
$\Omega_m$ also includes the contribution from massive neutrinos
besides the contributions from baryons and dark matter;
the dark energy density function $X(z)$ is defined as
\be
X(z) \equiv \frac{\rho_X(z)}{\rho_X(0)}.
\ee
Note that $\Omega_{\rm rad}=\Omega_m /(1+z_{\rm eq}) \ll \Omega_m$
(with $z_{\rm eq}$ denoting the redshift at matter-radiation equality),
thus the $\Omega_{\rm rad}$ term is usually omitted in dark energy studies at $z\ll 1000$,
since dark energy should only be important at the late times.

\subsection{SNe Ia data}

SNe Ia data give measurements of the luminosity distance $d_L(z)$
through that of the distance modulus of each SN:
\be
\label{eq:m-M}
\mu_0 \equiv m-M= 5 \log\left[\frac{d_L(z)}{\mathrm{Mpc}}\right]+25,
\ee
where $m$ and $M$ represent the apparent and absolute magnitude of an SN.
The luminosity distance $d_L(z)=(1+z)\, r(z)$, with the comoving distance $r(z)$ given by Eq. (\ref{eq:r(z)}).

Here we use the SNLS3 data set.
As mentioned above, based on different light-curve fitters,
three SNe sets of SNLS3 are given, including ``SALT2'', ``SiFTO'', and ``combined''.
As shown in \cite{WangWang},
the conclusion of evolution of $\beta$
is insensitive to the lightcurve fitter used to derive the SNLS3 sample.
So in this paper we just use the ``combined'' set.

In \cite{WangWang},
by considering three functional forms (linear case, quadratic case, and step function case),
the possible evolution of $\alpha$ and $\beta$ is explored.
It is found that, for the SNLS3 data,
$\alpha$ is still consistent with a constant,
but $\beta$ increases significantly with $z$.
It should be stressed that
this conclusion is insensitive to the functional form of $\alpha$ and $\beta$ assumed \cite{WangWang}.
So in this paper,
we just adopt a constant $\alpha$ and a linear $\beta(z) = \beta_{0} + \beta_{1} z$.
Now, the predicted magnitude of an SN becomes
\be
m_{\rm mod}=5 \log_{10}{\cal D}_L(z|\mbox{\bf p})
- \alpha (s-1) +\beta(z) {\cal C} + {\cal M},
\ee
where ${\cal D}_L(z|\mbox{\bf p})$ is the luminosity distance
multiplied by $H_0$ for a given set of cosmological parameters $\{ {\bf p} \}$,
$s$ is the stretch measure of the SN light curve shape,
and ${\cal C}$ is the color measure for the SN.
${\cal M}$ is a nuisance parameter representing some combination
of the absolute magnitude of a fiducial SN, $M$, and the Hubble constant, $H_0$.
In order to include host-galaxy information in the cosmological fits, the SNLS3 sample is splitted into two 
samples based on host-galaxy stellar mass at $10^{10} M_{\odot}$, and ${\cal M}$ is allowed to be 
different for the two samples~\cite{Conley11}. Therefore, for the SNLS3 sample, there are two values of ${\cal M}$, 
i.e., ${\cal M}_1$ and ${\cal M}_2$ (for more details, see the subsections $3.2$ and $5.8$ of~\cite{Conley11}).
The method of analytically marginalizing over ${\cal M}$ in this case is detailedly described in Appendix C of~\cite{Conley11}, 
and the corresponding code is public (now it is a part of CosmoMC). 
In this work, we follow the recipe of~\cite{Conley11}.

Since the time dilation part of the observed luminosity distance depends
on the total redshift $z_{\rm hel}$ (special relativistic plus cosmological),
we have
\be
{\cal D}_L(z|\mbox{\bf s}) = c^{-1}H_0 (1+z_{\rm hel}) r(z|\mbox{\bf s}),
\ee
where $z$ and $z_{\rm hel}$ are the CMB restframe and heliocentric redshifts of the SN.

For a set of $N$ SNe with correlated errors, we have \cite{Conley11}
\be
\label{eq:chi2_SN}
\chi^2=\Delta \mbox{\bf m}^T \cdot \mbox{\bf C}^{-1} \cdot \Delta\mbox{\bf m}
\ee
where $\Delta m \equiv m_B-m_{\rm mod}$ is a vector with $N$ components,
$m_B$ is the rest-frame peak B-band magnitude of the SN,
and $\mbox{\bf C}$ is the $N\times N$ covariance matrix of the SNe.

The total covariance matrix is \cite{Conley11}
\be
\mbox{\bf C}=\mbox{\bf D}_{\rm stat}+\mbox{\bf C}_{\rm stat}
+\mbox{\bf C}_{\rm sys},
\ee
with the diagonal part of the statistical uncertainty given by \cite{Conley11}
\ba
\mbox{\bf D}_{{\rm stat},ii}&=&\sigma^2_{m_B,i}+\sigma^2_{\rm int}
+ \sigma^2_{\rm lensing}+ \sigma^2_{{\rm host}\,{\rm correction}} \nonumber\\
&& + \left[\frac{5(1+z_i)}{z_i(1+z_i/2)\ln 10}\right]^2 \sigma^2_{z,i} \nonumber\\
&& +\alpha^2 \sigma^2_{s,i}+\beta(z_i)^2 \sigma^2_{{\cal C},i} \nonumber\\
&& + 2 \alpha C_{m_B s,i} - 2 \beta(z_i) C_{m_B {\cal C},i} \nonumber\\
&& -2\alpha \beta(z_i) C_{s {\cal C},i},
\ea
where $C_{m_B s,i}$, $C_{m_B {\cal C},i}$, and $C_{s {\cal C},i}$
are the covariances between $m_B$, $s$, and ${\cal C}$ for the $i$-th SN,
$\beta_i=\beta(z_i)$ is the value of $\beta$ for the $i$-th SN.
Note also that $\sigma^2_{z,i}$ includes a peculiar velocity residual of 0.0005 (i.e., 150$\,$km/s) added in quadrature \cite{Conley11}.
In this paper we just use the values of intrinsic scatter $\sigma_{\rm int}$ given in Table 4 of \cite{Conley11}.
These values are obtained by making $\chi^2/dof = 1$.
Varying $\sigma_{\rm int}$ could have a significant impact on parameter estimation, see \cite{Kim2011} for details.

We define $\mbox{\bf V} \equiv \mbox{\bf C}_{\rm stat} + \mbox{\bf C}_{\rm sys}$,
where $\mbox{\bf C}_{\rm stat}$ and $\mbox{\bf C}_{\rm sys}$
are the statistical and systematic covariance matrices, respectively.
After treating $\beta$ as a function of $z$,
$\mbox{\bf V}$ is given in the form:
\ba
\mbox{\bf V}_{ij}&=&V_{0,ij}+\alpha^2 V_{a,ij} + \beta_i\beta_j V_{b,ij} \nonumber\\
&& +\alpha V_{0a,ij} +\alpha V_{0a,ji} \nonumber\\
&& -\beta_j V_{0b,ij} -\beta_i V_{0b,ji} \nonumber\\
&& -\alpha \beta_j V_{ab,ij} - \alpha \beta_i V_{ab,ji}.
\ea
Here, $\beta_j=\beta(z_j)$ is the value of $\beta$ for the $j$-th SN;
the obvious difference from $\beta_0$ and $\beta_1$ in the form $\beta(z)=\beta_0+\beta_1 z$ should be noticed.
It must be stressed that, while $V_0$, $V_{a}$, $V_{b}$, and $V_{0a}$
are the same as the ``normal'' covariance matrices
given by the SNLS3 data archive, $V_{0b}$ and $V_{ab}$ are {\it not} the same as the ones given there.
This is because the original matrices of SNLS3 are produced by assuming that $\beta$ is a constant.
We have used the $V_{0b}$ and $V_{ab}$ matrices for the ``combined'' set
that are applicable when varying $\beta(z)$ (A.~Conley, private communication, 2013).

In \cite{WangWang}, it is found that
the flux-averaging of SNe \cite{Wang2000,WangPia04,Wang2005,WangChuPia2012}
may be helpful to reduce the effect of varying $\beta$.
It should be mentioned that
the results of flux-averaging depend on the choices of redshift cut-off $z_{\rm cut}$:
$\beta$ still increases with $z$ when all the SNe are flux-averaged,
and $\beta$ is consistent with being a constant
when only SNe at $z\ge 0.04$ are flux-averaged \cite{WangWang}.
Since the unknown systematic biases originate mostly from low $z$ SNe,
flux-averaging {\it all} SNe should lead to the least biased results.
Therefore, after applying the flux-averaging method,
the effect of varying $\beta$ is not completely removed.
For simplicity, we do not use the flux-averaging method in this paper,
and we will discuss the issue of flux-averaging in future work.

\subsection{CMB and GC data}
\label{sec:CMBGC}

For CMB data, we use the latest distance priors data
extracted from Planck first data release \cite{WangWangCMB}.

CMB give us the comoving distance to the photon-decoupling surface $r(z_*)$,
and the comoving sound horizon at photon-decoupling epoch $r_s(z_*)$.
Wang and Mukherjee \cite{WangPia07} showed that the CMB shift parameters
\ba
l_a &\equiv &\pi r(z_*)/r_s(z_*), \nonumber\\
R &\equiv &\sqrt{\Omega_m H_0^2} \,r(z_*)/c,
\ea
together with $\omega_b\equiv \Omega_b h^2$, provide an efficient summary
of CMB data as far as dark energy constraints go.
Replacing $\omega_b$ with $z_*$ gives identical constraints
when the CMB distance priors are combined with other data \cite{Wang08b}.
Using $\omega_b$, instead of $z_*$,
is more appropriate in a Markov Chain Monte Carlo (MCMC) analysis
in which $\omega_b$ is a base parameter.

The comoving sound horizon at redshift $z$ is given by
\ba
\label{eq:rs}
r_s(z)  &= & \int_0^{t} \frac{c_s\, dt'}{a}
=cH_0^{-1}\int_{z}^{\infty} dz'\,
\frac{c_s}{E(z')}, \nonumber\\
 &= & cH_0^{-1} \int_0^{a}
\frac{da'}{\sqrt{ 3(1+ \overline{R_b}\,a')\, {a'}^4 E^2(z')}},
\ea
where $a$ is the cosmic scale factor, $a =1/(1+z)$, and
$a^4 E^2(z)=\Omega_m (a+a_{\rm eq})+\Omega_k a^2 +\Omega_X X(z) a^4$,
with $a_{\rm eq}=\Omega_{\rm rad}/\Omega_m=1/(1+z_{\rm eq})$, and
$z_{\rm eq}=2.5\times 10^4 \Omega_m h^2 (T_{\rm cmb}/2.7\,{\rm K})^{-4}$.
The sound speed is $c_s=1/\sqrt{3(1+\overline{R_b}\,a)}$,
with $\overline{R_b}\,a=3\rho_b/(4\rho_\gamma)$,
$\overline{R_b}=31500\Omega_bh^2(T_{\rm cmb}/2.7\,{\rm K})^{-4}$.
We take $T_{\rm cmb}=2.7255\,{\rm K}$.

The redshift to the photon-decoupling surface, $z_*$, is given by the
fitting formula \cite{Hu96}:
\be
z_*=1048\, \left[1+ 0.00124 (\Omega_b h^2)^{-0.738}\right]\,
\left[1+g_1 (\Omega_m h^2)^{g_2} \right],
\ee
where
\ba
g_1 &= &\frac{0.0783\, (\Omega_b h^2)^{-0.238}}
{1+39.5\, (\Omega_b h^2)^{0.763}}, \\
g_2 &= &\frac{0.560}{1+21.1\, (\Omega_b h^2)^{1.81}}.
\ea
The redshift of the drag epoch $z_d$ is well approximated by \cite{EisenHu98}
\begin{equation}
z_d  =
 \frac{1291(\Omega_mh^2)^{0.251}}{1+0.659(\Omega_mh^2)^{0.828}}
\left[1+b_1(\Omega_bh^2)^{b2}\right],
\label{eq:zd}
\end{equation}
where
\begin{eqnarray}
  b_1 &= &0.313(\Omega_mh^2)^{-0.419}\left[1+0.607(\Omega_mh^2)^{0.674}\right],\\
  b_2 &= &0.238(\Omega_mh^2)^{0.223}.
\end{eqnarray}

Using the Planck+lensing+WP data,
the mean values and 1$\sigma$ errors of $\{ l_a, R, \omega_b\}$ are obtained \cite{WangWangCMB},
\ba
&&\langle l_a \rangle = 301.57, \sigma(l_a)=0.18, \nonumber\\
&&\langle R \rangle = 1.7407,  \sigma(R)=0.0094, \nonumber\\
&& \langle \omega_b \rangle = 0.02228, \sigma(\omega_b)=0.00030.
\label{eq:CMB_mean_planck}
\ea
Defining $p_1=l_a(z_*)$, $p_2=R(z_*)$, and $p_3=\omega_b$,
the normalized covariance matrix $\mbox{NormCov}_{CMB}(p_i,p_j)$ can be written as \cite{WangWangCMB}
\be
\left(
\begin{array}{ccc}
   1.0000  &    0.5250  &   -0.4235    \\
  0.5250  &     1.0000  &   -0.6925    \\
 -0.4235  &   -0.6925  &     1.0000    \\
\end{array}
\right).
\label{eq:normcov_planck}
\ee
Then, the covariance matrix for $(l_a, R, \omega_b)$ is given by
\be
\mbox{Cov}_{CMB}(p_i,p_j)=\sigma(p_i)\, \sigma(p_j) \,\mbox{NormCov}_{CMB}(p_i,p_j),
\label{eq:CMB_cov}
\ee
where $i,j=1,2,3$.
The CMB data are included in our analysis by adding
the following term to the $\chi^2$ function:
\be
\label{eq:chi2CMB}
\chi^2_{CMB}=\Delta p_i \left[ \mbox{Cov}^{-1}_{CMB}(p_i,p_j)\right]
\Delta p_j,
\hskip .2cm
\Delta p_i= p_i - p_i^{data},
\ee
where $p_i^{data}$ are the mean values from Eq. (\ref{eq:CMB_mean_planck}).

For GC data, we use the measurements of $H(z)r_s(z_d)/c$ and $D_A(z)/r_s(z_d)$
from the two-dimensional two-point correlation function
measured at $z=0.35$ \cite{ChuangWang12} and $z=0.57$ \cite{Chuang13},
where the angular diameter distance $D_A(z)= c H_0^{-1} r(z)/(1+z)$.
The $z=0.35$ measurement was made by Chuang and Wang \cite{ChuangWang12} using
a sample of the SDSS DR7 Luminous Red Galaxies (LRGs).
The $z=0.57$ measurement was made by Chuang et al. \cite{Chuang13} using the
CMASS galaxy sample from BOSS.

Using the two-dimensional two-point correlation function of SDSS DR7
in the scale range of 40--120$\,$Mpc/$h$,
Chuang and Wang \cite{ChuangWang12} found that
\ba
H(z=0.35)r_s(z_d)/c&=&0.0434  \pm  0.0018,  \nonumber \\
D_A(z=0.35)/r_s(z_d)&=& 6.60  \pm  0.26.
\label{eq:CW2}
\ea
where $r_s(z_d)$ is the sound horizon at the drag epoch given by
Eqs. (\ref{eq:rs}) and (\ref{eq:zd}).
In a similar analysis using the CMASS galaxy sample from BOSS,
Chuang et al. \cite{Chuang13} found that
\ba
H(z=0.57)r_s(z_d)/c&=&0.0454	\pm  0.0031,  \nonumber \\
D_A(z=0.57)/r_s(z_d)&=& 8.95	\pm  0.27.
\label{eq:C13}
\ea

GC data are included in our analysis by adding $\chi^2_{GC}=\chi^2_{GC1}+\chi^2_{GC2}$,
with $z_{GC1}=0.35$ and $z_{GC2}=0.57$, to $\chi^2$ of a given model.
Note that
\be
\label{eq:chi2bao}
\chi^2_{GCi}=\Delta q_i \left[ {\rm C}^{-1}_{GCi}(q_i,q_j)\right]
\Delta q_j,
\hskip .2cm
\Delta q_i= q_i - q_i^{data},
\ee
where $q_1=H(z_{GCi})r_s(z_d)/c$, $q_2=D_A(z_{GCi})/r_s(z_d)$, and $i=1,2$.
Based on Refs. \cite{ChuangWang12} and \cite{Chuang13}, we have
\begin{equation}
 {\rm C}_{GC1}=\left(
  \begin{array}{cc}
    0.00000324 & 0.0000282672 \\
    0.0000282672 & 0.0676 \\
  \end{array}
\right),
\end{equation}
\begin{equation}
 {\rm C}_{GC2}=\left(
  \begin{array}{cc}
    0.00000961 & 0.0004079538 \\
    0.0004079538 & 0.0729 \\
  \end{array}
\right).
\end{equation}

\section{Results}

As mentioned above, in this paper we consider three simplest models:
$\Lambda$CDM, $w$CDM, and CPL.
To explore the evolution of color-luminosity parameter $\beta$,
we study the case of constant $\alpha$ and linear $\beta(z) = \beta_{0} + \beta_{1} z$;
for comparison, the case of constant $\alpha$ and constant $\beta$ is also taken into account.

We perform an MCMC likelihood analysis \cite{COSMOMC}
to obtain ${\cal O}$($10^6$) samples for each set of results presented in this paper.
We assume flat priors for all the parameters, and allow ranges of the parameters wide enough
such that further increasing the allowed ranges has no impact on the results.
The chains typically have worst e-values
(the variance(mean)/mean(variance) of 1/2 chains) much smaller than $0.01$, indicating convergence.

In the following,
we will discuss the results given by the SNe-only and the SNe+CMB+GC data, respectively.

\subsection{SNe-only cases}
\label{sec:SNeonly}

In this subsection, we discuss the results given by the SNe-only data.
Notice that the Hubble constant $h$ has been marginalized during the $\chi^2$ fitting process of SNe Ia,
so we only need to consider six free parameters,
including $\alpha$, $\beta_0$, $\beta_1$, $\Omega_{m}$,
$w_0$, and $w_1$ [two parameters for the equation of state $w(z)$].
In Table \ref{table1}, we list the fitting results for various constant $\beta$ and linear $\beta(z)$ cases,
where only SNe data are used.
The most obvious feature of this table is that varying $\beta$ can significantly improve the fitting results.
This conclusion is insensitive to the dark energy models:
for all the models,
adding a parameter of $\beta$ can reduce the best-fit values of $\chi^2$ by $\sim$ 36.
Based on the Wilk's theorem, 36 units of $\chi^2$ is equivalent to a Gaussian fluctuation of 6$\sigma$.
This means that for all the models, the result of $\beta_1=0$ is ruled out.
It must be stressed that the evolution of $\beta$ is not a special feature only found in the SNLS3 data.
In \cite{Mohlabeng}, Mohlabeng and Ralston showed that, for the Union2.1 SN data,
adding a parameter of $\beta$ can reduce the best-fit values of $\chi^2$ by $\sim$ 50.
Therefore, it is very necessary and important to consider $\beta$'s evolution in the cosmology-fits.

\begin{table*}\tiny
\caption{Fitting results for various constant $\beta$ and linear $\beta(z)$ cases,
where only SNe data are used.}
\label{table1}
\begin{tabular}{ccccccccc}
\hline\hline &\multicolumn{2}{c}{$\Lambda$CDM} &&\multicolumn{2}{c}{$w$CDM} && \multicolumn{2}{c}{CPL} \\
           \cline{2-3}\cline{5-6}\cline{8-9}
Parameters  & Const $\beta$ & Linear $\beta(z)$ && Const $\beta$ & Linear $\beta(z)$ && Const $\beta$ & Linear $\beta(z)$\\ \hline
$\alpha$           & $1.425^{+0.109}_{-0.103}$
                   & $1.410^{+0.106}_{-0.094}$&
                   & $1.427^{+0.108}_{-0.101}$
                   & $1.410^{+0.103}_{-0.092}$&
                   & $1.427^{+0.106}_{-0.106}$
                   & $1.415^{+0.096}_{-0.097}$\\

$\beta_0$          & $3.259^{+0.110}_{-0.108}$
                   & $1.457^{+0.370}_{-0.376}$&
                   & $3.256^{+0.114}_{-0.102}$
                   & $1.439^{+0.398}_{-0.336}$&
                   & $3.265^{+0.104}_{-0.109}$
                   & $1.499^{+0.300}_{-0.453}$  \\

$\beta_1$          & N/A
                   & $5.061^{+1.064}_{-1.027}$&
                   & N/A
                   & $5.112^{+0.970}_{-1.074}$&
                   & N/A
                   & $4.939^{+1.256}_{-0.796}$ \\

$\Omega_{m}$      & $0.226^{+0.040}_{-0.036}$
                   & $0.280^{+0.052}_{-0.052}$&
                   & $0.163^{+0.100}_{-0.147}$
                   & $0.135^{+0.215}_{-0.009}$&
                   & $0.320^{+0.055}_{-0.310}$
                   & $0.252^{+0.137}_{-0.242}$\\

$w_0$              & N/A
                   & N/A&
                   & $-0.858^{+0.219}_{-0.224}$
                   & $-0.630^{+0.058}_{-0.268}$&
                   & $-0.778^{+0.235}_{-0.268}$
                   & $-0.667^{+0.254}_{-0.240}$\\

$w_1$              & N/A
                   & N/A&
                   & N/A
                   & N/A&
                   & $-3.619^{+4.370}_{-1.380}$
                   & $-2.260^{+2.609}_{-2.739}$\\

\hline $\chi^{2}_{min}$  & 420.075  & 385.203 & & 419.658 & 383.591 && 419.054&383.144  \\
\hline
\end{tabular}
\end{table*}

In addition, we feel that it may be necessary to report the fit results of all the parameters, including the nuisance 
parameter ${\cal M}$, in order for our work to be reproducible for the reader. Therefore, here we give the best-fit 
values of  ${\cal M}_1$ and ${\cal M}_2$ for the SNe-only cases.
For the constant $\beta$ case:
${\cal M}_1 = 0.01327$ and ${\cal M}_2 = -0.03708$ for the $\Lambda$CDM model;
${\cal M}_1 = 0.01503$ and ${\cal M}_2 = -0.03815$ for the $w$CDM model;
${\cal M}_1 = 0.01839$ and ${\cal M}_2 = -0.03682$ for the CPL model.
For the linear $\beta(z)$ case:
${\cal M}_1 = -0.00343$ and ${\cal M}_2 = -0.05650$ for the $\Lambda$CDM model;
${\cal M}_1 = -0.001348$ and ${\cal M}_2 = -0.06815$ for the $w$CDM model;
${\cal M}_1 = 0.00703$ and ${\cal M}_2 = -0.05333$ for the CPL model.

Let us discuss the SNe-only cases with more details.

\begin{itemize}
 \item $\Lambda$CDM model
\end{itemize}

\begin{figure}
\includegraphics[scale=0.3, angle=0]{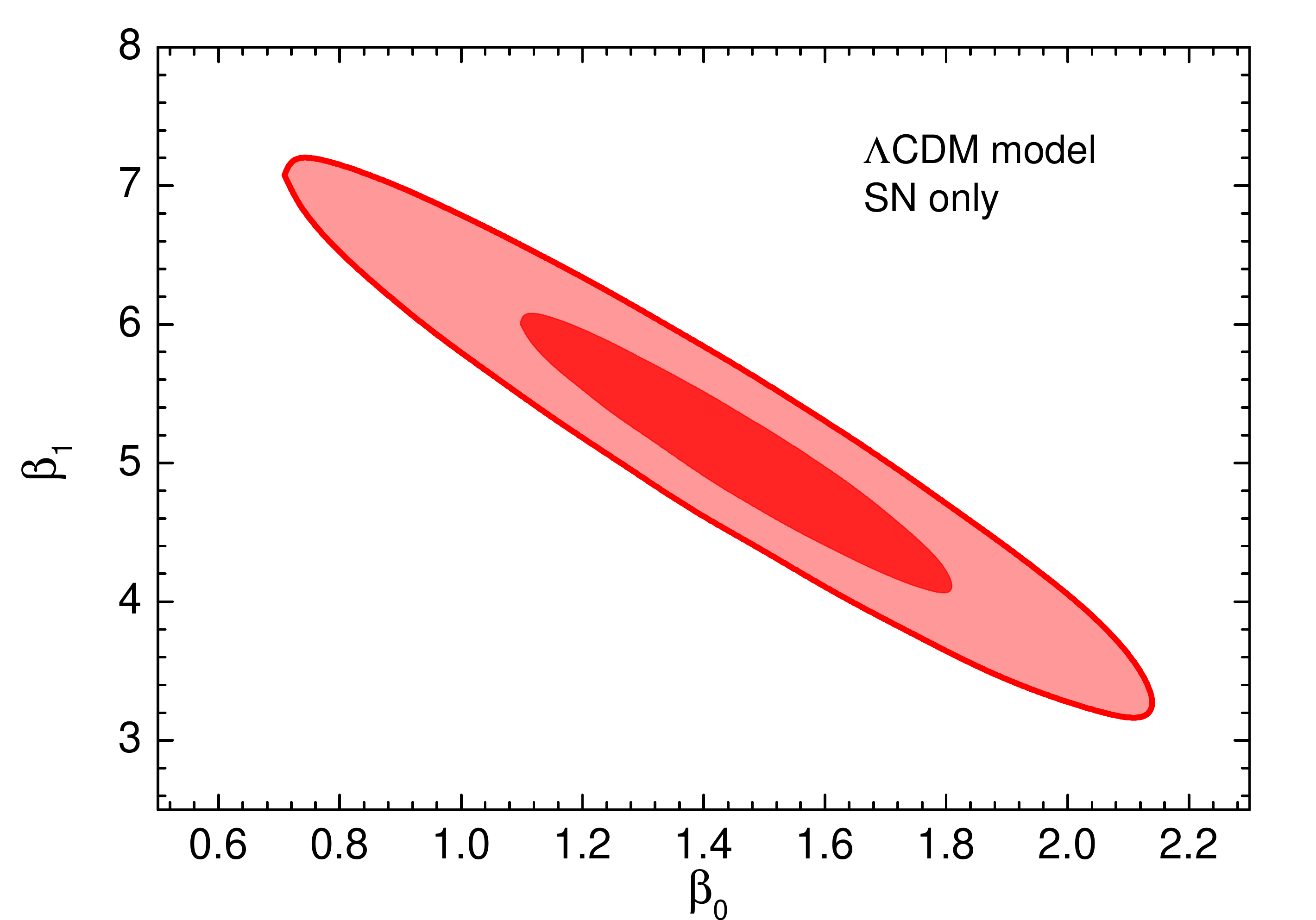}
\includegraphics[scale=0.3, angle=0]{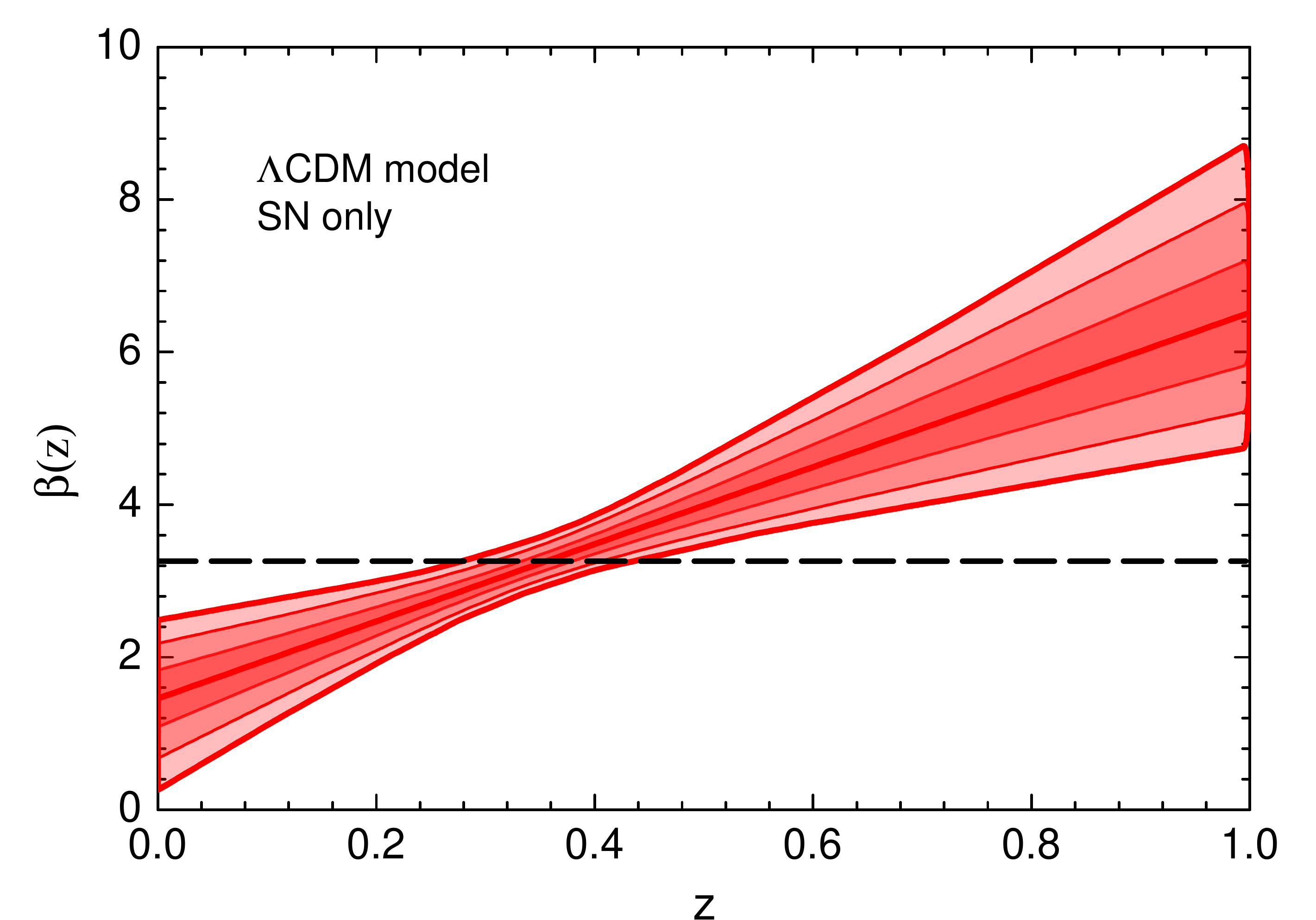}
\caption{\label{fig1}\footnotesize%
The joint $68\%$ and $95\%$ confidence contours for $\{\beta_{0},\beta_{1}\}$ (top panel),
and the $68\%$, $95\%$, and $97\%$ confidence constraints for $\beta(z)$ (bottom panel),
given by the SNe-only data, for the $\Lambda$CDM model.
For comparison,
the best-fit result of constant $\beta$ case is also shown on the bottom panel.}
\end{figure}

Firstly, we discuss the results of the $\Lambda$CDM model.
In Fig. \ref{fig1}, using SNe-only data,
we plot the joint $68\%$ and $95\%$ confidence contours for $\{\beta_{0},\beta_{1}\}$ (top panel),
and the $68\%$, $95\%$, and $97\%$ confidence constraints for $\beta(z)$ (bottom panel),
for the linear $\beta(z)$ case.
For comparison, we also show the best-fit result of constant $\beta$ case on the bottom panel.
The top panel shows that $\beta_1 > 0$ at a high confidence level (CL).
Here we adopt $\chi^2 \equiv \chi^2_{min} + i^2$ with $i=1$, 2, and 3,
corresponding to 1--3 units of Gaussian CL $\sigma$.
In addition, there is a clear degeneracy between $\beta_0$ and $\beta_1$,
which may be due to the kinematic fact of fitting a linear function.
The bottom panel shows that $\beta(z)$ rapidly increases with $z$.
Moreover, comparing with the best-fit result of constant $\beta$ case,
one can see that $\beta$ deviates from a constant at 6$\sigma$ CL.
It needs to be pointed out that the evolutionary behaviors of $\beta(z)$ depends on the SN samples used.
In \cite{Mohlabeng}, Mohlabeng and Ralston found that, for the Union2.1 SN data, $\beta(z)$ decreases with $z$.
This is similar to the case of Pan-STARRS1 SN data \cite{Scolnic}.
It is of great interest to study why different SN data give different evolutionary behaviors of $\beta(z)$,
and some numerical simulation studies may be required to solve this problem.
We will study this issue in future work.

\begin{figure}
\psfig{file=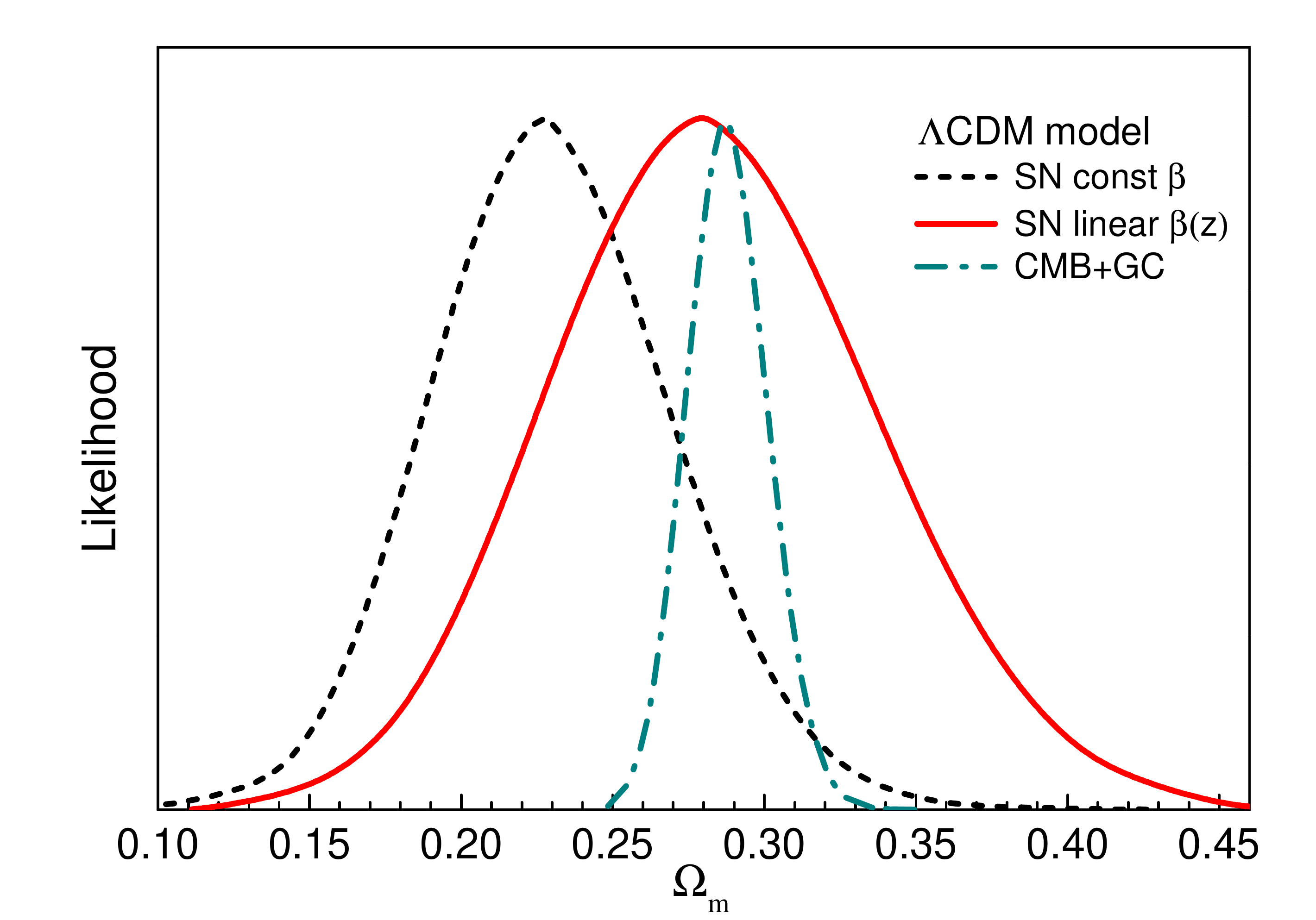,width=3.5in}\\
\caption{\label{fig2}\footnotesize%
The 1D marginalized probability distributions of $\Omega_m$,
given by the SNe-only data, for the $\Lambda$CDM model.
Both the results of constant $\beta$ and linear $\beta(z)$ cases are presented.
The corresponding results given by the CMB+GC data are also shown for comparison.
}
\end{figure}

Now, we study the effects of varying $\beta$ on parameter estimation of $\Lambda$CDM model.
In Fig. \ref{fig2}, using SNe-only data,
we plot the 1D marginalized probability distributions of $\Omega_m$
for both the constant $\beta$ and linear $\beta(z)$ cases.
We find that varying $\beta$ yields a larger $\Omega_m$:
the best-fit result for the constant $\beta$ case is $\Omega_m=0.226$,
while the best-fit result for the linear $\beta(z)$ case is $\Omega_m=0.280$.
To make a direct comparison,
we also plot the 1D distribution of $\Omega_m$ given by the CMB+GC data,
and find that the best-fit result for this case is $\Omega_m=0.287$.
Therefore, the result of linear $\beta(z)$ case
is much closer to that given by the CMB+GC data,
compared to the case of treating $\beta$ as a constant.
This means that varying $\beta$ is very helpful to reduce the tension
between SNe and other cosmological observations.
It should be mentioned that, for different SN data,
the effects of varying $\beta$ on parameter estimation are different.
For example, for the Union2.1 data, varying $\beta$ yields a smaller $\Omega_m$:
the best-fit value of $\Omega_m$ is revised from $\Omega_m=0.29$ to $\Omega_m=0.26$ \cite{Mohlabeng}.
For this case, there is no significant tension between SNe and other cosmological observations.
This shows that there still exists significant disagreement between different SN samples.

\begin{figure}
\psfig{file=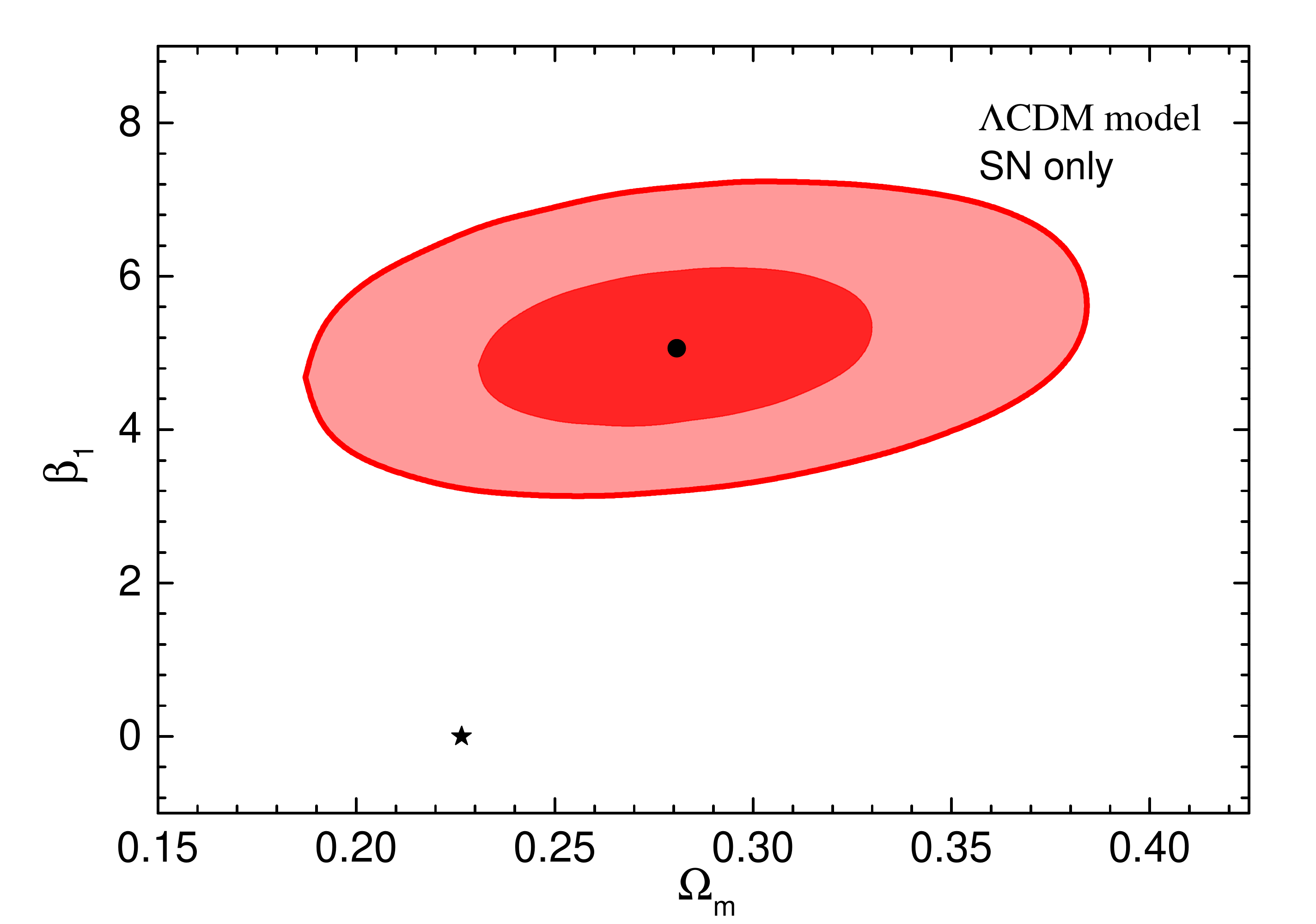,width=3.5in}\\
\caption{\label{fig3}\footnotesize%
The joint $68\%$ and $95\%$ confidence contours for $\{\Omega_m,\beta_1\}$,
given by the SNe-only data, for the $\Lambda$CDM model.
To make a direct comparison,
we also plot the best-fit points for the constant $\beta$ (star symbol) and the linear $\beta(z)$ (round point) cases.
}
\end{figure}

In Fig. \ref{fig3}, using the SNe-only data,
we plot the joint $68\%$ and $95\%$ confidence contours for $\{\Omega_m,\beta_1\}$, for the $\Lambda$CDM model.
As shown in Table \ref{table1}, $\Omega_m$ is related to the value of $\beta_1$.
To make a direct comparison,
we also plot the best-fit point (star symbol) of the constant $\beta$ case,
which corresponds to $\chi^2_{min} = 420.075$.
Compared with the best-fit point (round point, corresponding to $\chi^2_{min} = 385.203$) of the linear $\beta(z)$ case,
a constant $\beta$ will enlarge the best-fit value of $\chi^2$ by $34.872$, which is equivalent to a Gaussian fluctuation of 5.9$\sigma$.
This means that a constant $\beta$ is ruled out at 5.9$\sigma$ CL.

\begin{itemize}
 \item $w$CDM and CPL models
\end{itemize}

\begin{figure}
\includegraphics[scale=0.3, angle=0]{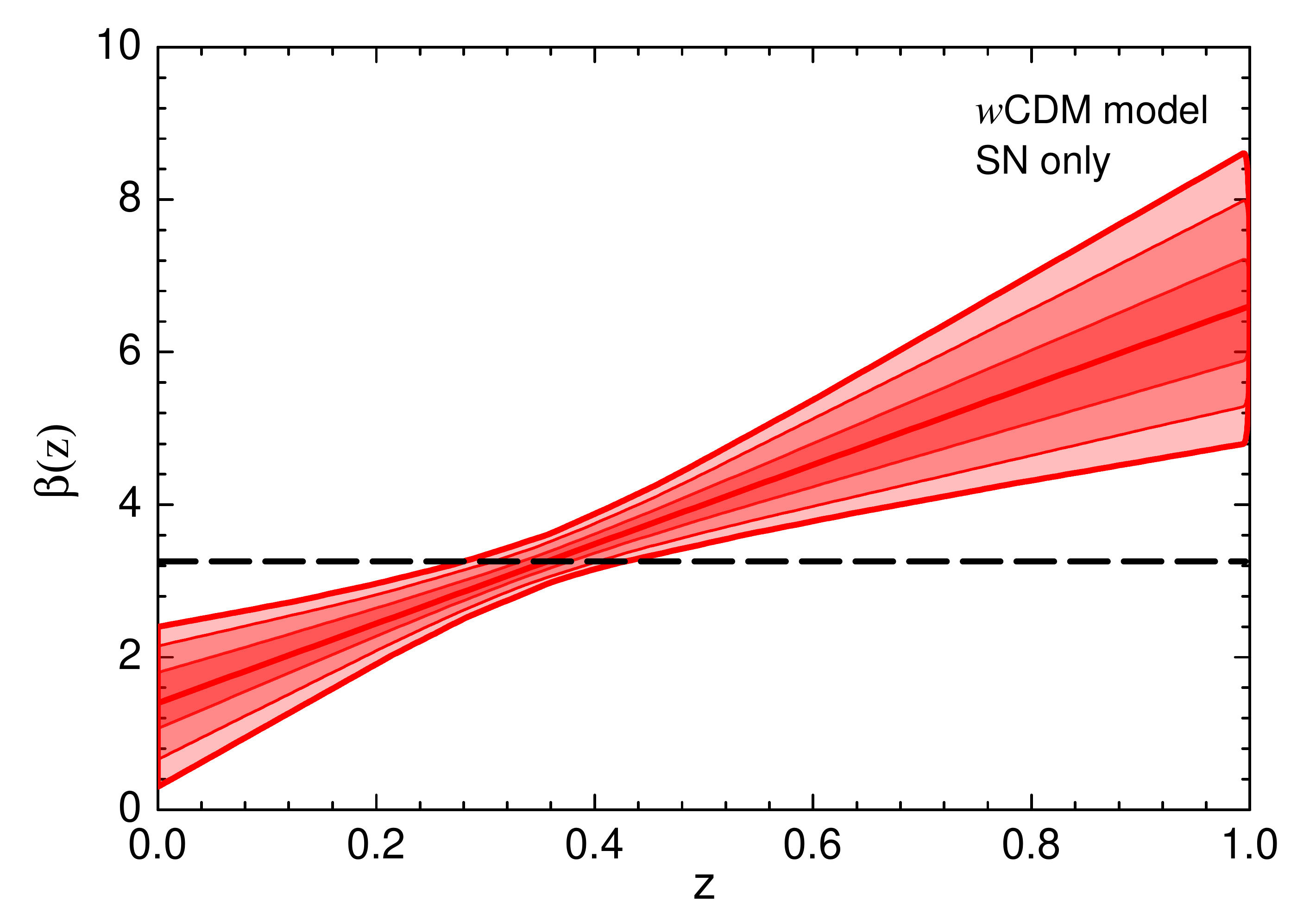}
\includegraphics[scale=0.3, angle=0]{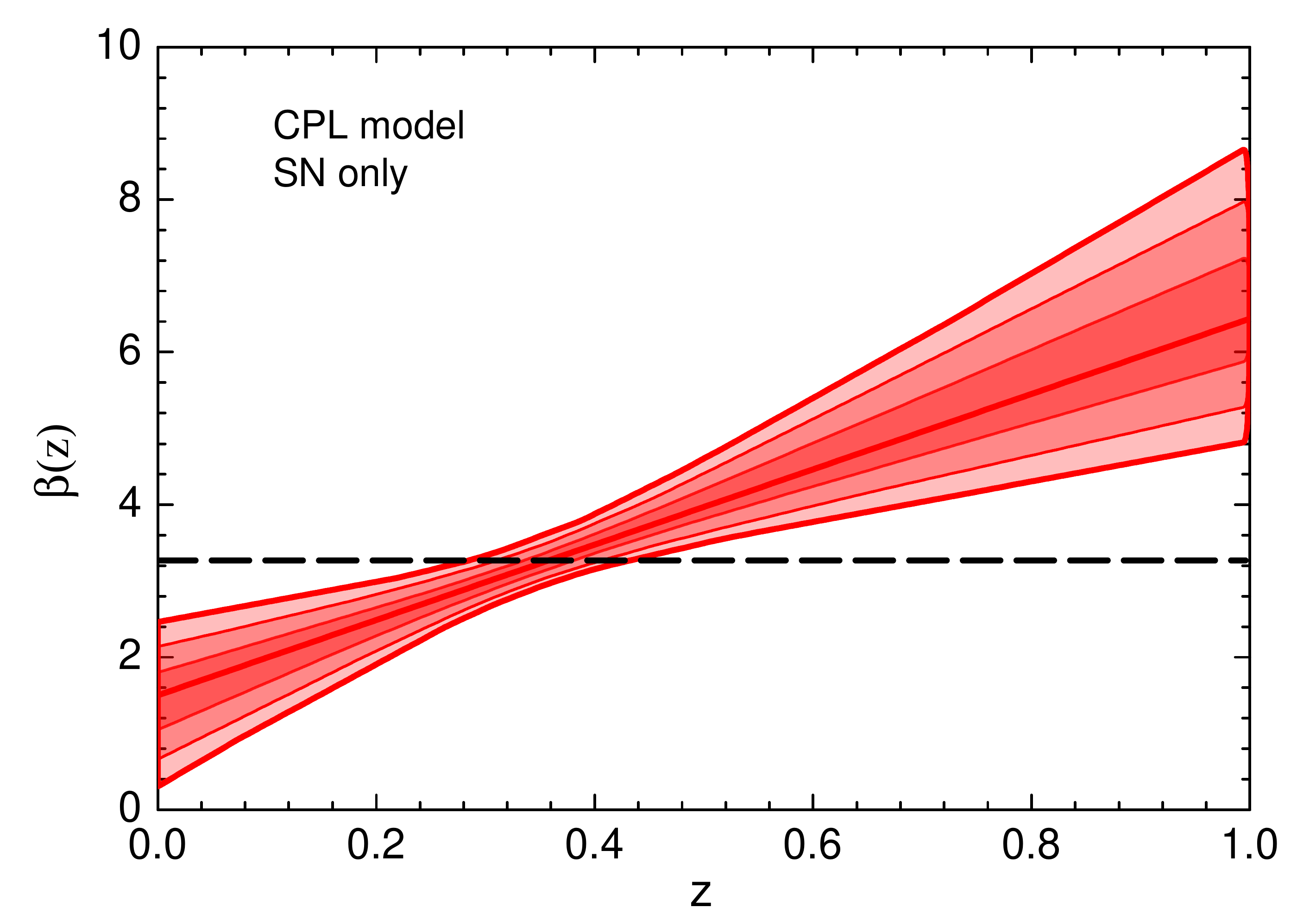}
\caption{\label{fig4}\footnotesize%
The $68\%$, $95\%$, and $97\%$ confidence constraints for $\beta(z)$, given by the SNe-only data,
for the $w$CDM model (top panel) and the CPL model (bottom panel).
For comparison,
the best-fit results of constant $\beta$ cases are also shown.}
\end{figure}

Next, we discuss the results of the $w$CDM model and the CPL model.
In Fig. \ref{fig4}, using SNe-only data,
we plot the $68\%$, $95\%$, and $97\%$ confidence constraints for $\beta(z)$,
for the $w$CDM model (top panel) and the CPL model (bottom panel).
It is clear that for both the $w$CDM model and the CPL model,
$\beta(z)$ rapidly increase with $z$.
Moreover, comparing with the best-fit results of constant $\beta$ case,
one can see that $\beta$ deviates from a constant at 6$\sigma$ CL.
Since the results of Fig. \ref{fig1} and Fig. \ref{fig4} are very close,
we can conclude that the evolution of $\beta$ is insensitive to the models considered.
Based on Table \ref{table1},
one can see that, for both the $w$CDM model and the CPL model,
varying $\beta$ yields a smaller $\Omega_m$,
compared to the cases of assuming a constant $\beta$.
This result is different from that of the $\Lambda$CDM model,
and is also different from the results given by the SNe+CMB+GC data (see next subsection).
This may be due to using SNe data alone still has difficulty
to break the degeneracy between $\Omega_m$ and $w$.

\subsection{SNe+CMB+GC cases}
\label{sec:Combindata}

In this subsection, we discuss the results given by the SNe+CMB+GC data.
It should be mentioned that, in order to use the Planck distance priors data,
three new model parameters, including $h$, $\omega_b$, and $\Omega_{k}$, must be added.
In Table \ref{table2}, we list the fitting results for various constant $\beta$ and linear $\beta(z)$ cases,
where the SNe+CMB+GC data are used.
Again, we find that varying $\beta$ can significantly improve the fitting results.
For all the dark energy models,
adding a parameter of $\beta$ can reduce the best-fit values of $\chi^2$ by $\sim$ 36.
This means that after considering the CMB and the GC data,
the result of $\beta_1=0$ is still ruled out for all the models.
This shows the importance of considering the evolution of $\beta$ in the cosmology-fits.

\begin{table*}\tiny
\caption{Fitting results for various constant $\beta$ and linear $\beta(z)$ cases,
where the SNe+CMB+GC data are used.}
\label{table2}
\begin{tabular}{ccccccccc}
\hline\hline &\multicolumn{2}{c}{$\Lambda$CDM} &&\multicolumn{2}{c}{$w$CDM} && \multicolumn{2}{c}{CPL} \\
           \cline{2-3}\cline{5-6}\cline{8-9}
Parameters  & Const $\beta$ & Linear $\beta(z)$ && Const $\beta$ & Linear $\beta(z)$ && Const $\beta$ & Linear $\beta(z)$\\ \hline
$\alpha$           & $1.429^{+0.099}_{-0.111}$
                   & $1.421^{+0.093}_{-0.100}$&
                   & $1.433^{+0.095}_{-0.108}$
                   & $1.425^{+0.086}_{-0.105}$&
                   & $1.438^{+0.090}_{-0.103}$
                   & $1.421^{+0.079}_{-0.093}$\\

$\beta_0$          & $3.249^{+0.109}_{-0.106}$
                   & $1.400^{+0.394}_{-0.326}$&
                   & $3.253^{+0.109}_{-0.099}$
                   & $1.493^{+0.300}_{-0.420}$&
                   & $3.269^{+0.099}_{-0.104}$
                   & $1.478^{+0.258}_{-0.389}$  \\

$\beta_1$          & N/A
                   & $5.208^{+0.890}_{-1.074}
$&
                   & N/A
                   & $4.960^{+1.089}_{-0.798}$&
                   & N/A
                   & $5.012^{+1.100}_{-0.735}$ \\

$\Omega_{m}$      & $0.281^{+0.013}_{-0.010}$
                   & $0.287^{+0.011}_{-0.013}$&
                   & $0.270^{+0.014}_{-0.013}$
                   & $0.286^{+0.013}_{-0.016}$&
                   & $0.275^{+0.012}_{-0.013}$
                   & $0.280^{+0.013}_{-0.015}$\\

$h$                & $0.704^{+0.013}_{-0.014}$
                   & $0.698^{+0.015}_{-0.012}$&
                   & $0.719^{+0.016}_{-0.018}$
                   & $0.698^{+0.021}_{-0.016}$&
                   & $0.714^{+0.019}_{-0.015}$
                   & $0.706^{+0.020}_{-0.016}$\\

$\omega_b$        & $0.02233^{+0.00028}_{-0.00030}$
                   & $0.02226^{+0.00030}_{-0.00026}$&
                   & $0.02229^{+0.00028}_{-0.00028}$
                   & $0.02235^{+0.00022}_{-0.00034}$&
                   & $0.02228^{+0.00027}_{-0.00028}$
                   & $0.02229^{+0.00023}_{-0.00028}$\\

$\Omega_{k}$      & $0.0031^{+0.0035}_{-0.0035}$
                   & $0.0024^{+0.0037}_{-0.0033}$&
                   & $0.0009^{+0.0031}_{-0.0045}$
                   & $0.0019^{+0.0051}_{-0.0036}$&
                   & $-0.0076^{+0.0053}_{-0.0033}$
                   & $-0.0093^{+0.0054}_{-0.0029}$\\

$w_0$              & N/A
                   & N/A &
                   & $-1.091^{+0.064}_{-0.085}$
                   & $-1.002^{+0.078}_{-0.075}$&
                   & $-0.783^{+0.162}_{-0.226}$
                   & $-0.619^{+0.190}_{-0.209}$\\

$w_1$              & N/A
                   & N/A &
                   & N/A
                   & N/A &
                   & $-2.180^{+1.424}_{-1.097}$
                   & $-3.059^{+1.610}_{-1.394}$\\

\hline $\chi^{2}_{min}$  & 423.922  & 387.077 & & 422.296 & 387.041 && 420.022&383.826  \\
\hline
\end{tabular}
\end{table*}

We also give the best-fit values of ${\cal M}_1$ and ${\cal M}_2$ for the SNe+CMB+GC cases.
For the constant $\beta$ case:
${\cal M}_1 = 0.006862$ and ${\cal M}_2 = -0.048252$ for the $\Lambda$CDM model;
${\cal M}_1 = 0.01078$ and ${\cal M}_2 = -0.03941$ for the $w$CDM model;
${\cal M}_1 = 0.01878$ and ${\cal M}_2 = -0.03796$ for the CPL model.
For the linear $\beta(z)$ case:
${\cal M}_1 = -0.003554$ and ${\cal M}_2 = -0.05799$ for the $\Lambda$CDM model;
${\cal M}_1 = -0.003131$ and ${\cal M}_2 = -0.05785$ for the $w$CDM model;
${\cal M}_1 = 0.008096$ and ${\cal M}_2 = -0.05369$ for the CPL model.

Let us discuss the effects of varying $\beta$ on various dark energy models in detail.

\begin{itemize}
 \item $\Lambda$CDM model
\end{itemize}

\begin{figure}
\includegraphics[scale=0.3, angle=0]{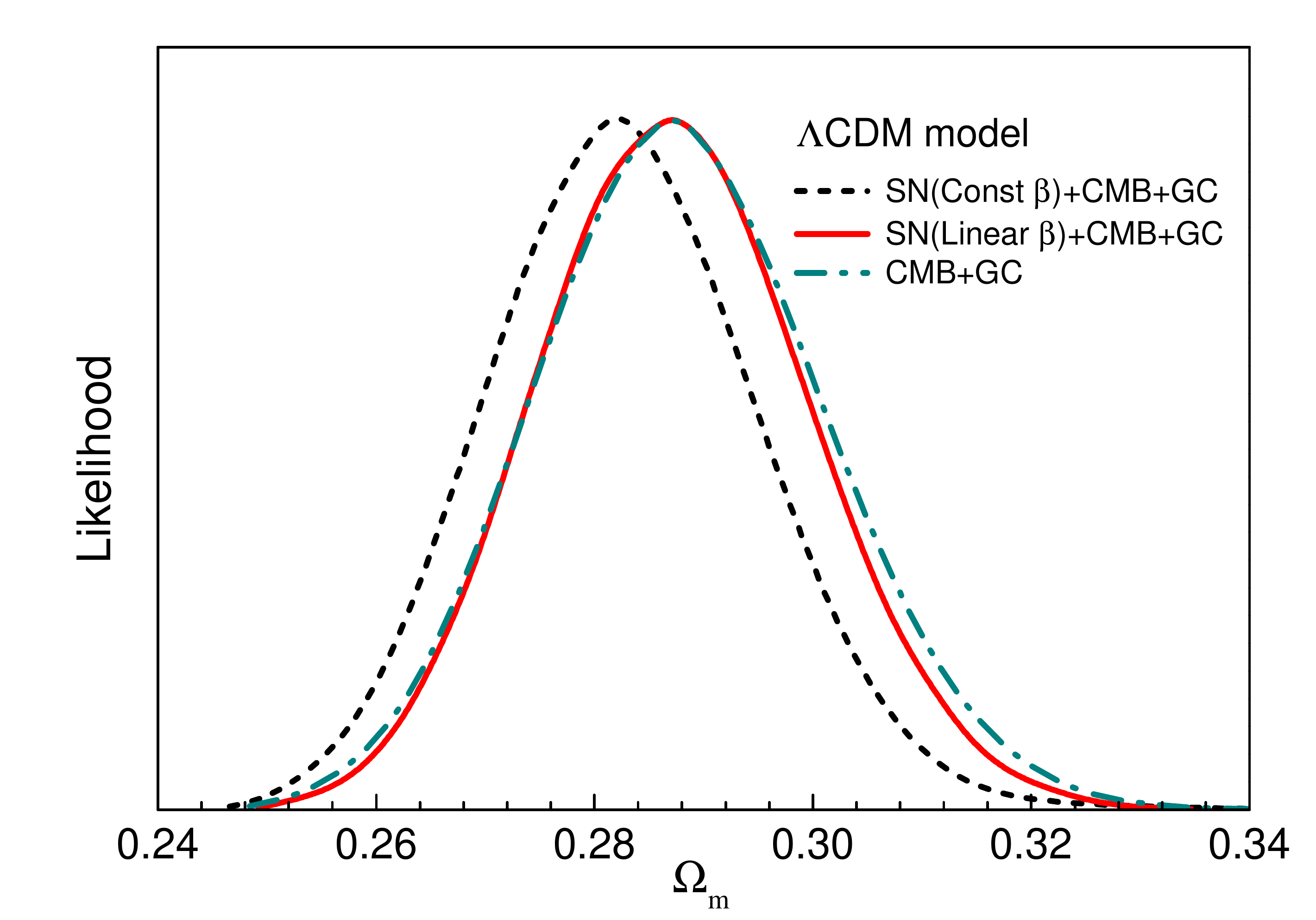}
\includegraphics[scale=0.3, angle=0]{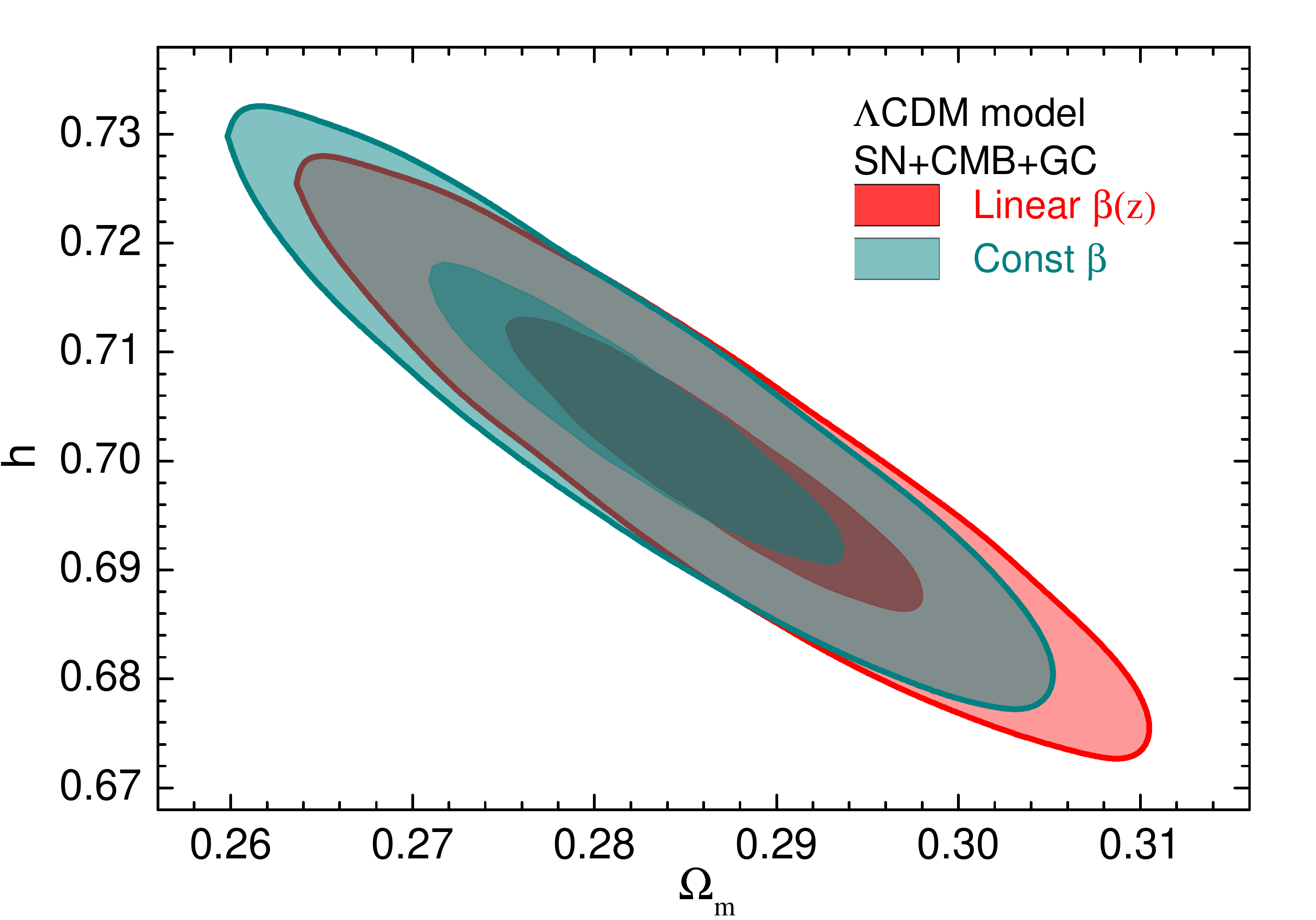}
\caption{\label{fig5}\footnotesize%
The 1D marginalized probability distribution of $\Omega_m$ (top panel),
and the joint $68\%$ and $95\%$ confidence contours for $\{\Omega_m,h\}$ (bottom panel),
given by the SNe+CMB+GC data, for the $\Lambda$CDM model.
Both the results of constant $\beta$ and linear $\beta(z)$ cases are shown.
The corresponding results given by the CMB+GC data are also shown for comparison.}
\end{figure}

Firstly, we start from the $\Lambda$CDM model.
In Fig. \ref{fig5}, using the SNe+CMB+GC data,
we plot the 1D marginalized probability distributions of $\Omega_m$ (top panel),
and the joint $68\%$ and $95\%$ confidence contours for $\{\Omega_m,h\}$ (bottom panel), for the $\Lambda$CDM model.
From the top panel,
we see that varying $\beta$ yields a larger $\Omega_m$:
the best-fit value of $\Omega_m$ for the constant $\beta$ case is $0.281$,
while best-fit value of $\Omega_m$ for the linear $\beta(z)$ case is $0.287$.
To make a direct comparison,
we also plot the 1D distribution of $\Omega_m$ given by the CMB+GC data.
It is clear that the 1D distribution of $\Omega_m$ for the linear $\beta(z)$ case
is closer to that given by the CMB+GC data.
So we can conclude that varying $\beta$ is very helpful to reduce the tension
between SNe and other cosmological observations.
This conclusion is consistent with that of Fig. \ref{fig2}.
From the bottom panel, we see that varying $\beta$ will also yield a smaller $h$:
the best-fit value of $h$ for the constant $\beta$ case is $0.704$,
while the best-fit value of $h$ for the linear $\beta(z)$ case is $0.698$.
In addition, it is clear that $\Omega_m$ and $h$ are anti-correlated.

\begin{figure}
\psfig{file=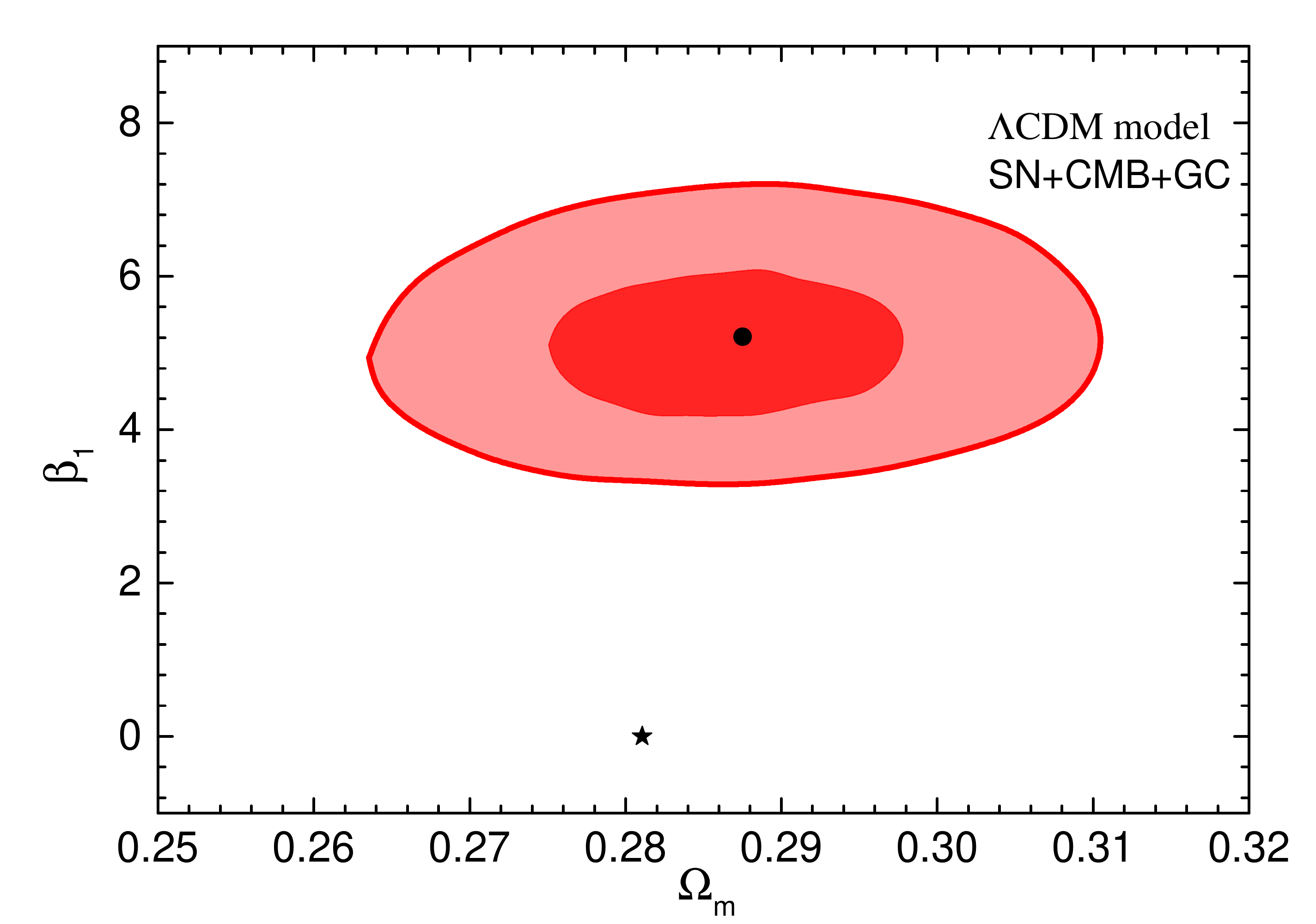,width=3.5in}\\
\caption{\label{fig6}\footnotesize%
The joint $68\%$ and $95\%$ confidence contours for $\{\Omega_m,\beta_1\}$,
given by the SNe+CMB+GC data, for the $\Lambda$CDM model.
To make a direct comparison,
we also plot the best-fit points for the constant $\beta$ (star symbol) and the linear $\beta(z)$ (round point) cases.
}
\end{figure}

In Fig. \ref{fig6}, using the SNe+CMB+GC data,
we plot the joint $68\%$ and $95\%$ confidence contours for $\{\Omega_m,\beta_1\}$, for the $\Lambda$CDM model.
As shown in Table \ref{table2}, $\Omega_m$ is related to the value of $\beta_1$.
To make a direct comparison,
we also plot the best-fit point (star symbol) of the constant $\beta$ case,
which corresponds to $\chi^2_{min} = 423.922$.
Compared with the best-fit point (round point, corresponding to $\chi^2_{min} = 387.077$) of the linear $\beta(z)$ case,
a constant $\beta$ will enlarge the best-fit values of $\chi^2$ by $36.845$, which is equivalent to a Gaussian fluctuation of 6.1$\sigma$.
This means that a constant $\beta$ is ruled out at 6.1$\sigma$ CL.
Notice that using SNe-only data can only rule out a constant $\beta$ at 5.9$\sigma$ CL,
so we can conclude that adding the CMB and GC data will strengthen the conclusion of $\beta_1 \neq 0$.

\begin{itemize}
 \item $w$CDM model
\end{itemize}

\begin{figure}
\includegraphics[scale=0.3, angle=0]{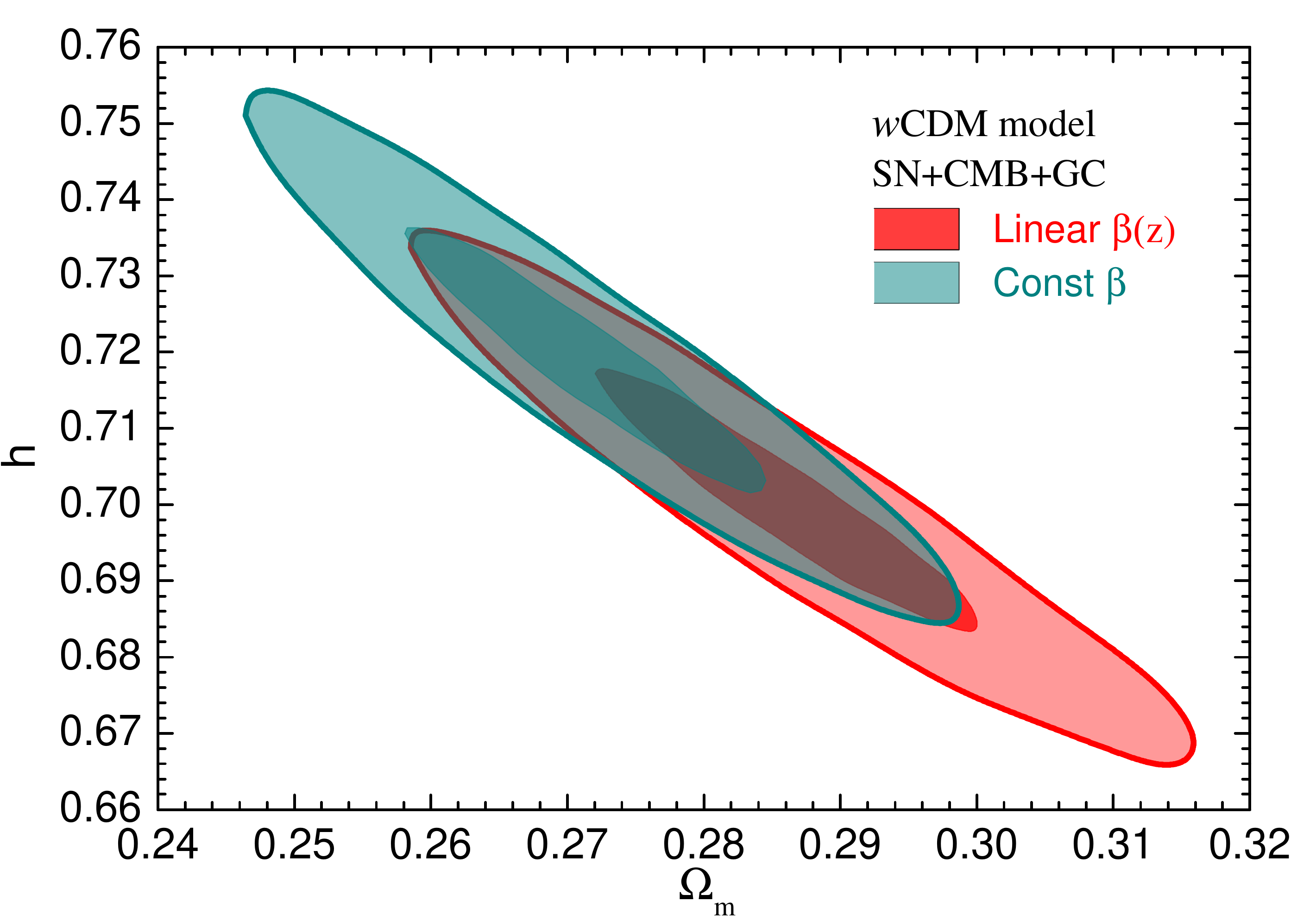}
\includegraphics[scale=0.3, angle=0]{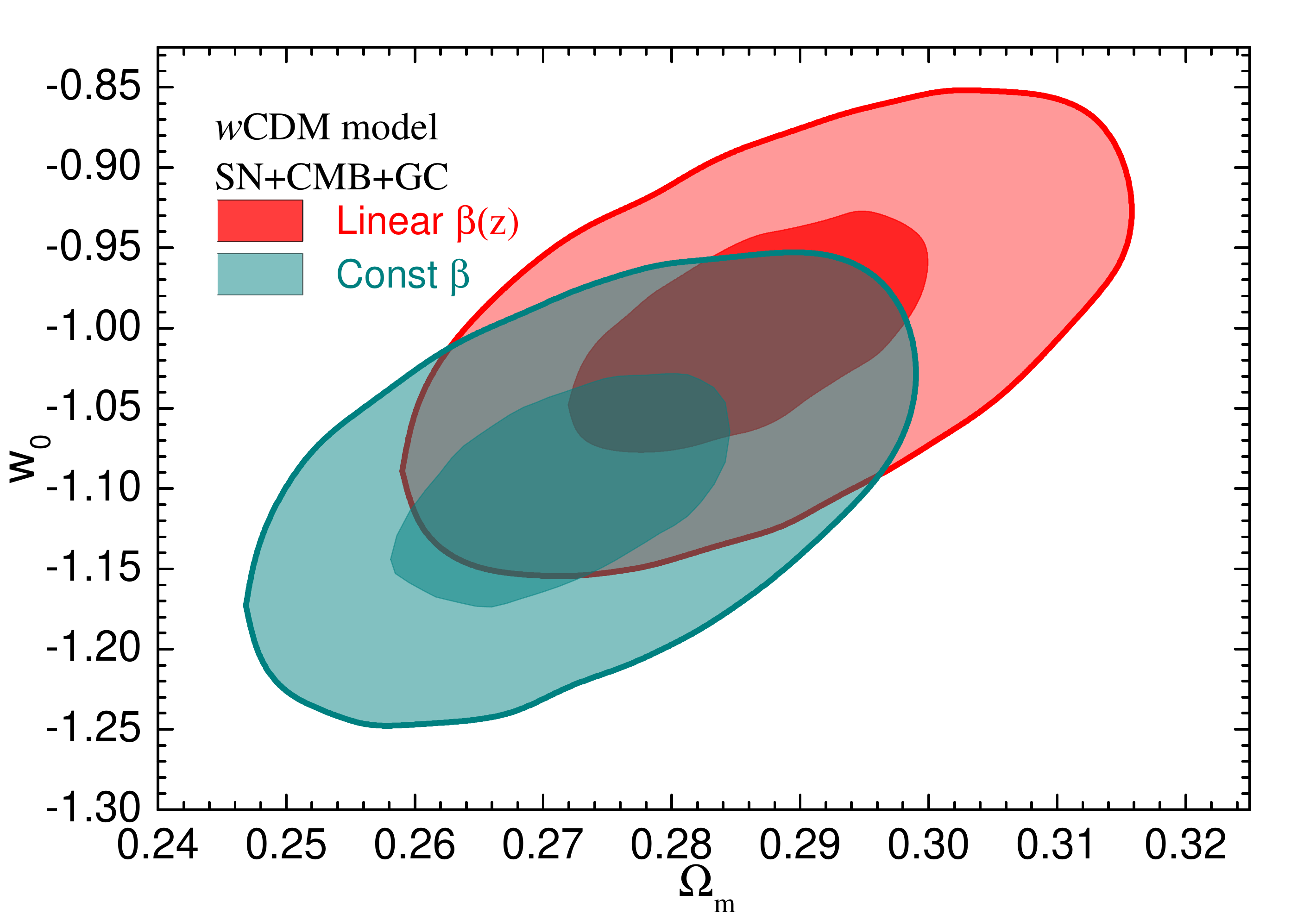}
\caption{\label{fig7}\footnotesize%
The joint $68\%$ and $95\%$ confidence contours
for $\{\Omega_m,h\}$ (top panel) and $\{\Omega_m,w_0\}$ (bottom panel),
given by the SNe+CMB+GC data, for the $w$CDM model.
Both the results of constant $\beta$ and linear $\beta(z)$ cases are shown for comparison.}
\end{figure}

Then, we turn to the $w$CDM model.
In Fig. \ref{fig7}, using the SNe+CMB+GC data,
we plot the joint $68\%$ and $95\%$ confidence contours
for $\{\Omega_m,h\}$ (top panel) and $\{\Omega_m,w_0\}$ (bottom panel), for the $w$CDM model.
From the top panel,
we see that varying $\beta$ yields a larger $\Omega_m$ and a smaller $h$:
the best-fit results for the constant $\beta$ case are $\Omega_m=0.270$ and $h=0.719$,
while best-fit results for the linear $\beta(z)$ case are $\Omega_m=0.286$ and $h=0.698$.
In addition, $\Omega_m$ and $h$ are anti-correlated.
This is consistent with the case of the $\Lambda$CDM model.
The bottom panel shows that varying $\beta$ will also yield a larger $w_0$:
the best-fit value of $w_0$ for the constant $\beta$ case is $-1.091$,
while the best-fit value of $w_0$ for the linear $\beta(z)$ case is $-1.002$.
Notice that after considering the evolution of $\beta$,
the results of the $w$CDM model are closer to that of the $\Lambda$CDM model.
In addition, $\Omega_m$ and $w_0$ are also in positive correlation.

\begin{itemize}
 \item CPL model
\end{itemize}

\begin{figure}
\includegraphics[scale=0.3, angle=0]{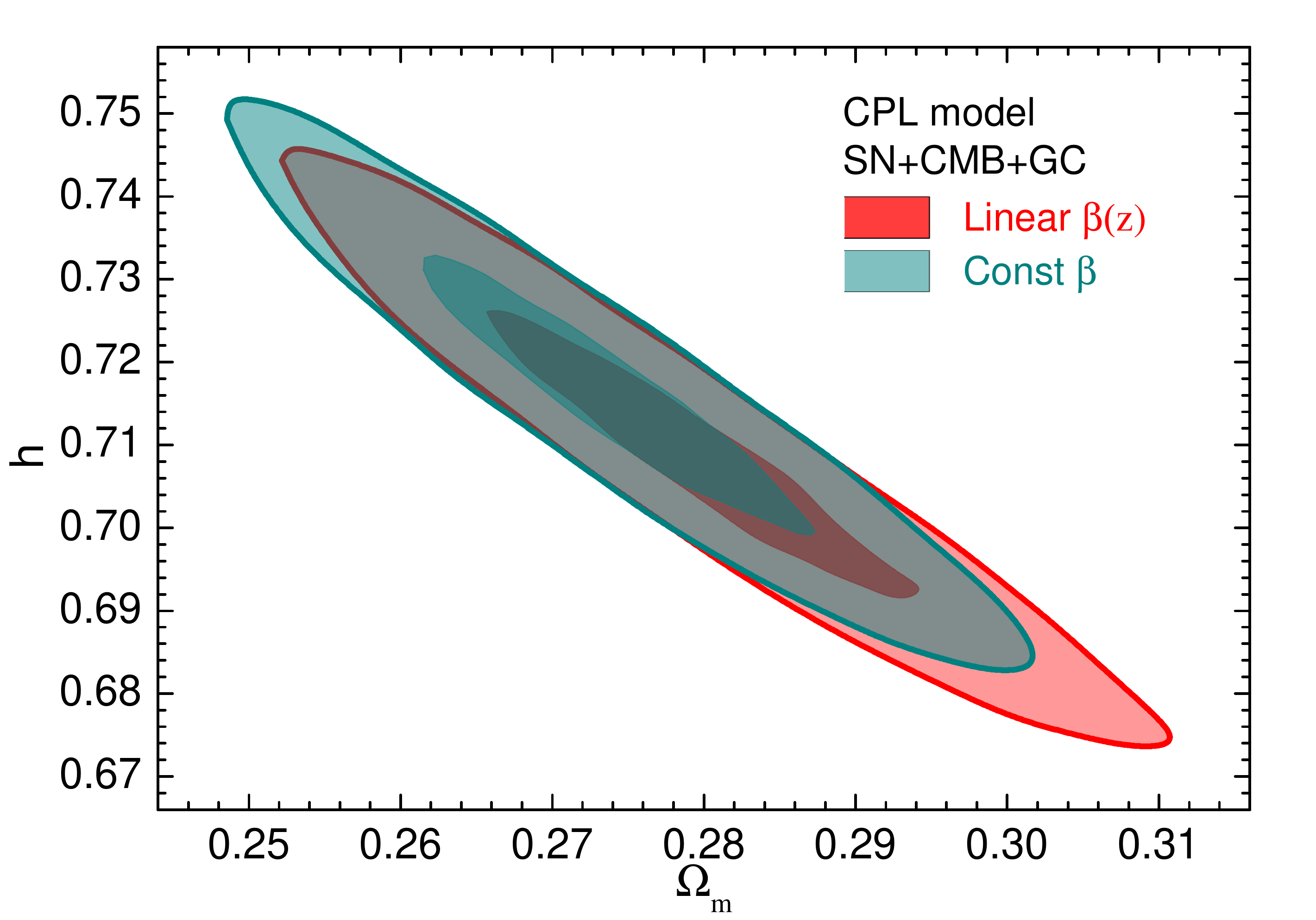}
\includegraphics[scale=0.3, angle=0]{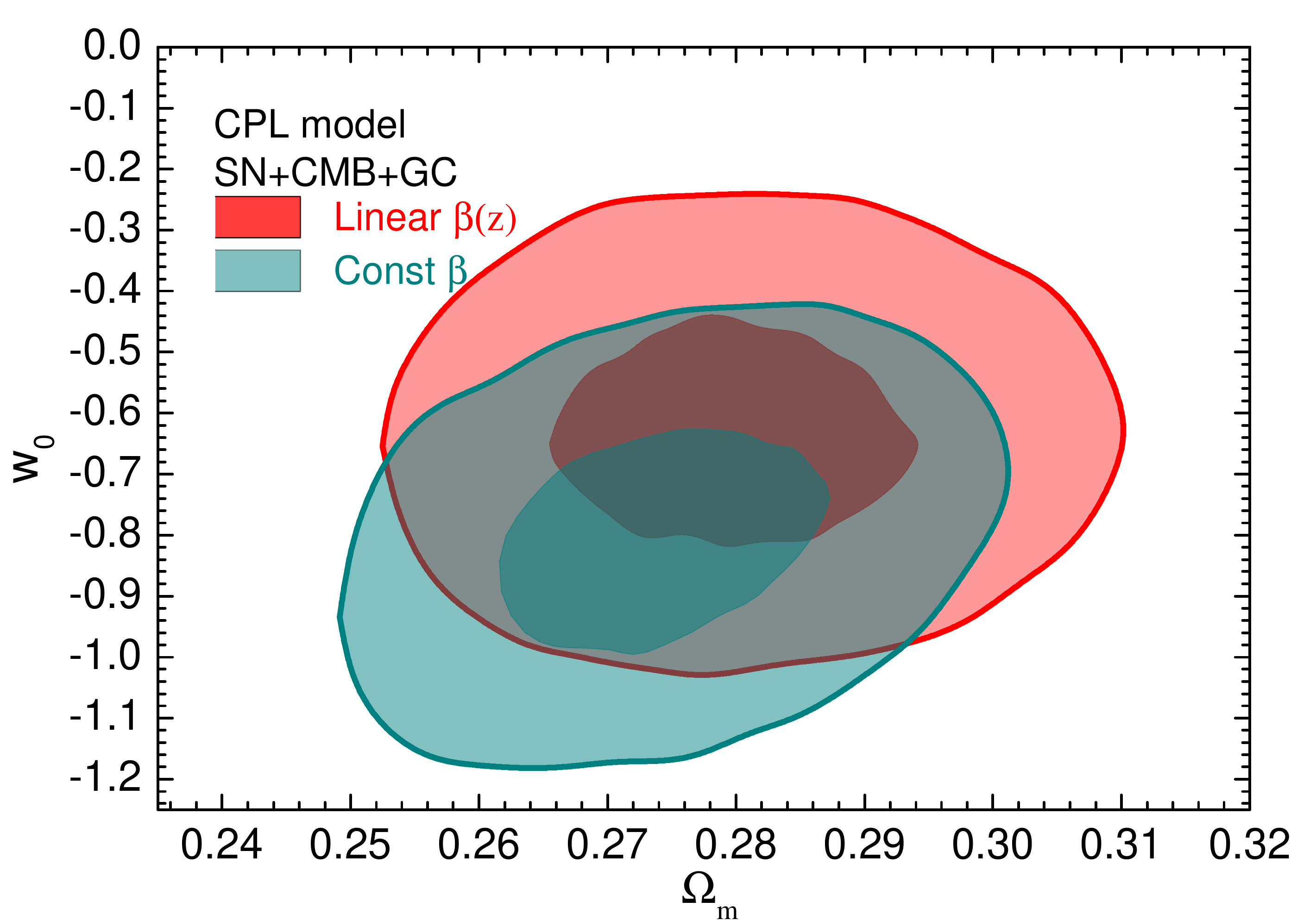}
\caption{\label{fig8}\footnotesize%
The joint $68\%$ and $95\%$ confidence contours
for $\{\Omega_m,h\}$ (top panel) and $\{\Omega_m,w_0\}$ (bottom panel),
given by the SNe+CMB+GC data, for the CPL model.
Both the results of constant $\beta$ and linear $\beta(z)$ cases are shown for comparison.}
\end{figure}

Next, we discuss the CPL model.
In Fig. \ref{fig8}, using the SNe+CMB+GC data,
we plot the joint $68\%$ and $95\%$ confidence contours
for $\{\Omega_m,h\}$ (top panel) and $\{\Omega_m,w_0\}$ (bottom panel), for the CPL model.
Again, we see from the top panel that
varying $\beta$ yields a larger $\Omega_m$ and a smaller $h$:
the best-fit results for the constant $\beta$ case are $\Omega_m=0.275$ and $h=0.714$,
while the best-fit results for the linear $\beta(z)$ case are $\Omega_m=0.280$ and $h=0.706$.
In addition, $\Omega_m$ and $h$ are also anti-correlated.
The bottom panel shows that varying $\beta$ will also yield a larger $w_0$:
the best-fit value of $w_0$ for the constant $\beta$ case is $-0.783$,
while the best-fit value of $w_0$ for the linear $\beta(z)$ case is $-0.619$.
These results are consistent with the cases of the $\Lambda$CDM model and the $w$CDM model.
To make a direct comparison,
we also study the CPL model using the CMB+GC data,
and find that the best-fit results for this case are $\Omega_m=0.282$, $h=0.709$ and $w_0=-0.712$.
It is clear that the fitting results for the linear $\beta(z)$ case
are much closer to that given by the CMB+GC data,
compared to the case of treating $\beta$ as a constant.
This indicates that the conclusion of Figs. \ref{fig2} and \ref{fig5}
is insensitive to the dark energy models considered.

\begin{figure}
\includegraphics[scale=0.3, angle=0]{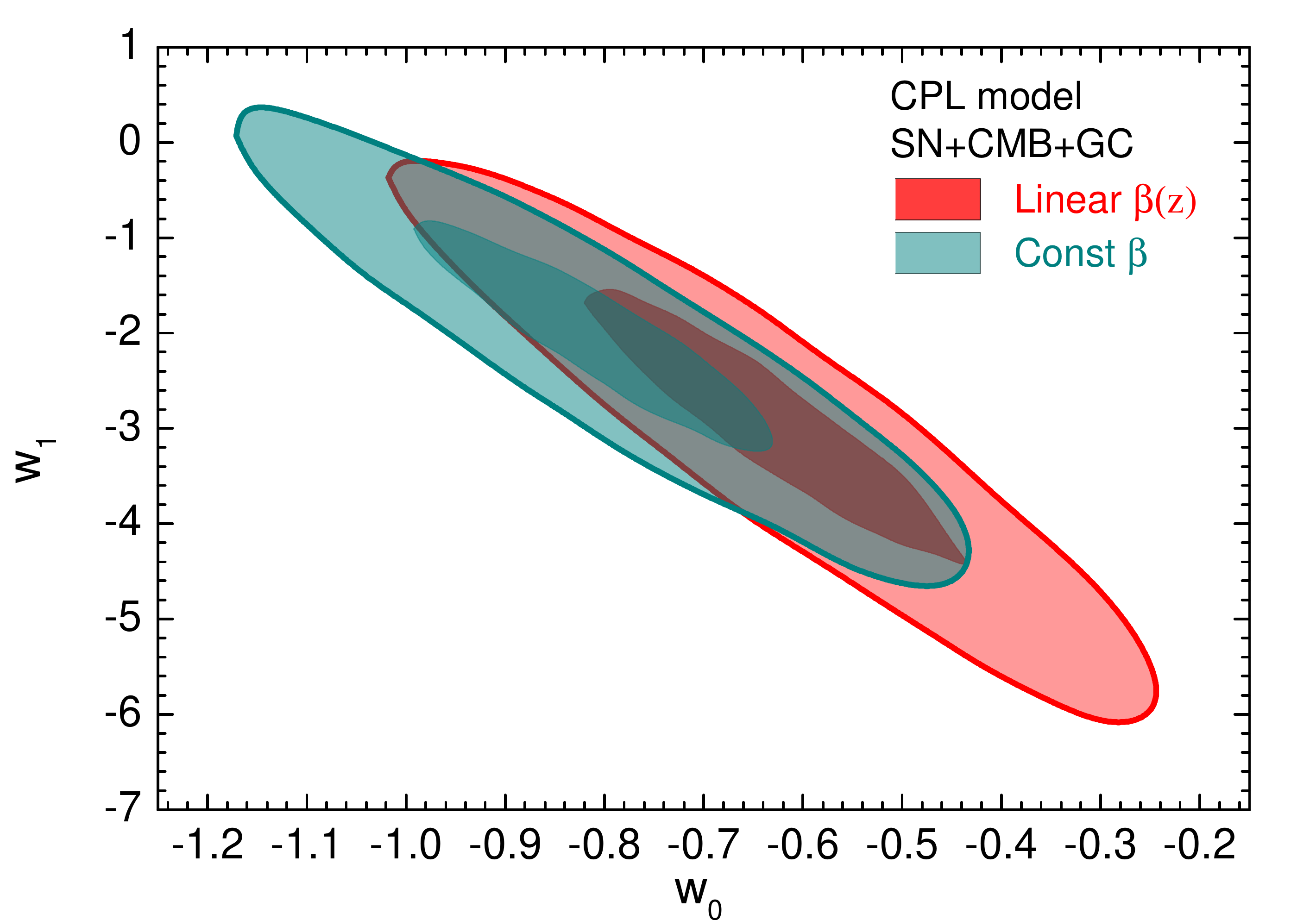}
\includegraphics[scale=0.3, angle=0]{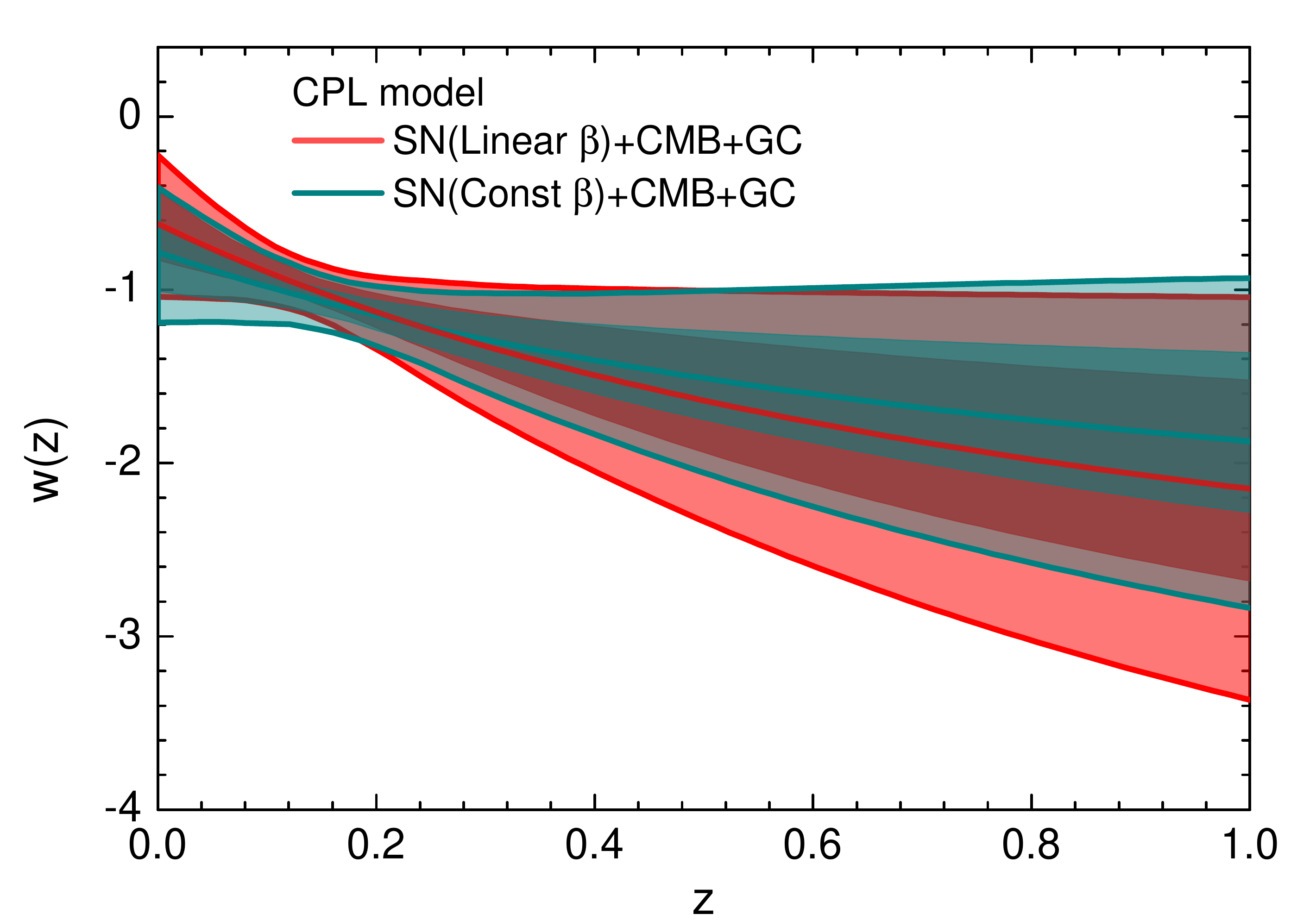}
\caption{\label{fig9}\footnotesize%
The joint $68\%$ and $95\%$ confidence contours for $\{w_0,w_1\}$ (top panel),
and the $68\%$ and $95\%$ confidence constraints for $w(z)$ (bottom panel),
given by the SNe+CMB+GC data, for the CPL model.
Both the results of constant $\beta$ and linear $\beta(z)$ cases are shown for comparison.}
\end{figure}

Finally, we discuss the effects of varying $\beta$ on the equation of state (EOS) $w(z)$ of the CPL model.
In Fig. \ref{fig9}, using the SNe+CMB+GC data,
we plot the joint $68\%$ and $95\%$ confidence contours for $\{w_0,w_1\}$ (top panel),
and the $68\%$ and $95\%$ confidence constraints for $w(z)$ (bottom panel), for the CPL model.
The top panel shows that varying $\beta$ yields a larger $w_0$ and a smaller $w_1$,
while $w_0$ and $w_1$ are anti-correlated.
The bottom panel shows that after considering the evolution of $\beta$,
EOS $w(z)$ of the CPL model will decrease faster with redshift $z$.

\section{Discussion and Summary}

Along with the rapid progress of SN cosmology, more and more SNe Ia have been discovered,
and the systematic errors of SNe Ia have drawn more and more attentions.
One of the most important systematic uncertainties for SNe Ia is the potential SN evolution.
The hints for the evolution of $\beta$ have been found in \cite{Astier06,Kowalski08,Kessler09,Guy10,Marriner11,Sullivan11},
but these papers explored $\beta$'s evolution using bin-by-bin fits,
which were very difficult to make definitive statements because of the correlations between different bins.
In \cite{Mohlabeng}, Mohlabeng and Ralston firstly used a linear parametrization $\beta(z) = \beta_0 + \beta_1 z$ to study the Union2.1 sample,
and found that $\beta$ deviates from a constant at 7$\sigma$ confidence levels.
Moreover, they proved that using a linear parametrization form can obtain better results than using bin-by-bin methods.
Wang and Wang \cite{WangWang} studied the case of SNLS3 data using three functional forms,
and also found strong evidence for the redshift-evolution of $\beta$.

It is clear that a time-varying $\beta$ will have significant impact on parameter estimation.
So in this paper, by adopting a constant $\alpha$ and a linear $\beta(z) = \beta_{0} + \beta_{1} z$,
we have further explored the evolution of $\beta$ and its effects on parameter estimation.
To perform the cosmology-fits,
we have considered three simplest dark energy models: $\Lambda$CDM, $w$CDM, and CPL.
In addition to the SNLS3 SN data,
we have also taken into account the Planck distance priors data,
as well as the latest GC data extracted from SDSS DR7 and BOSS.

We further confirm the redshift-evolution of $\beta$ for the SNLS3 data:
For all the models, adding a parameter of $\beta$ can reduce $\chi^2_{min}$ by $\sim$ 36,
indicating that $\beta_1 = 0$ is ruled out at 6$\sigma$ CL.
In other words, $\beta$ deviates from a constant at 6$\sigma$ CL.
This conclusion is insensitive to the dark energy models considered and the SN data used,
showing the importance of considering the evolution of $\beta$ in the cosmology-fits.

Furthermore,
it is found that varying $\beta$ can significantly change the fitting results of various cosmological parameters:
using the SNLS3 data alone,
varying $\beta$ yields a larger $\Omega_m$ for the $\Lambda$CDM model;
using the SNLS3+CMB+GC data,
varying $\beta$ yields a larger $\Omega_m$ and a smaller $h$ for all the models.
For the $w$CDM model, varying $\beta$ will also yield a larger $w_0$;
for the CPL model, varying $\beta$ yields a larger $w_0$ and a smaller $w_1$.
Moreover, we find that these results are much closer to those given by the CMB+GC data,
compared to the cases of treating $\beta$ as a constant.
This indicates that considering the evolution of $\beta$ is very helpful
for reducing the tension between supernova and other cosmological observations.

In this paper, only three simplest dark energy models are considered.
It is of interest to study the effects of varying $\beta$ on parameter estimation in other dark energy models.
In addition, some other factors, such as the evolution of $\sigma_{int}$ \cite{Kim2011},
may also cause the systematic uncertainties of SNe Ia.
These issues will be studied in future works.

\begin{acknowledgments}
We thank the referee for valuable suggestions, which help us to improve this work significantly.
We also thank Dr. Daniel Scolnic for helpful discussions.
We are grateful to Dr. Alex Conley for providing us with the SNLS3 covariance matrices that
allow redshift-dependent $\beta$. We acknowledge the use of CosmoMC.
This work is supported by the National Natural Science Foundation of
China (Grants No.~10975032 and No.~11175042) and by the National Ministry
of Education of China (Grants No.~NCET-09-0276 and No.~N120505003).
\end{acknowledgments}

\end{document}